\documentclass[aps,prd,reprint,superscriptaddress,nofootinbib,floatfix,longbibliography]{revtex4-1}
\usepackage[dvipsnames]{xcolor}
\usepackage{amssymb,amsmath,amsfonts,bm,slashed,soul,ragged2e,graphicx,epstopdf,hyperref,array,gensymb}
\usepackage[utf8]{inputenc}
\usepackage{colortbl,color,caption,subcaption}
\enlargethispage{\baselineskip}
\definecolor{steelblue}{RGB}{25,25,112}
\definecolor{dullblue}{rgb}{0,0.298,0.49}
\definecolor{darkred}{rgb}{0.545,0,0}
\definecolor{darkorange}{RGB}{222,132,69}
\definecolor{darkgreen}{RGB}{126,171,85}
\definecolor{blue2}{cmyk}{1, 0.1, 0.1, 0}
\hypersetup{colorlinks,linkcolor={darkred},citecolor={blue},urlcolor={dullblue}}
\DeclareCaptionJustification{justified}{\justifying}
\everymath{\displaystyle}

\usepackage{titlesec}

\usepackage{comment}

\newcommand{\beq}{\begin{equation}}
\newcommand{\eeq}{\end{equation}}
\newcommand{\bea}{\begin{eqnarray}}
\newcommand{\eea}{\end{eqnarray}}

\newcommand{\gsim}{\lower.7ex\hbox{$\;\stackrel{\textstyle>}{\sim}\;$}}
\newcommand{\lsim}{\lower.7ex\hbox{$\;\stackrel{\textstyle<}{\sim}\;$}}

\newcommand{\be}{\begin{equation}}
\newcommand{\ee}{\end{equation}}
\newcommand{\ba}{\begin{eqnarray}}
\newcommand{\ea}{\end{eqnarray}}

\def \d {{\rm d}}
\begin{document}

\title{Illuminating Black Hole Shadow with Dark Matter Annihilation}

\author{Yifan Chen}
\altaffiliation{Corresponding author: yifan.chen@nanograv.org}
\affiliation{Center of Gravity, Niels Bohr Institute, Blegdamsvej 17, 2100 Copenhagen, Denmark}

\author{Ran Ding}
\altaffiliation{Corresponding author: dingran@mail.nankai.edu.cn}
\affiliation{School of Physics and Optoelectronics Engineering, Anhui University, Hefei 230601,China}

\author{Yuxin Liu}
\affiliation{University of Chinese Academy of Sciences, Beijing 100190, China}
\affiliation{International Center for Theoretical Physics Asia-Pacific, Beijing/Hangzhou, China}
\affiliation{Department of Physics and Astronomy,
University of Utah, Salt Lake City, Utah 84112, USA}

\author{Yosuke Mizuno}
\altaffiliation{Corresponding author: mizuno@sjtu.edu.cn}
\affiliation{Tsung-Dao Lee Institute, Shanghai Jiao Tong University, 201210, Shanghai, China}
\affiliation{School of Physics \& Astronomy, Shanghai Jiao-Tong University, 200240, Shanghai, China}
\affiliation{Key Laboratory for Particle Astrophysics and Cosmology (MOE) and Shanghai Key Laboratory for Particle Physics and Cosmology, Shanghai Jiao-Tong University, 200240, Shanghai, China}

\author{Jing Shu}
\altaffiliation{Corresponding author: jshu@pku.edu.cn}
\affiliation{School of Physics and State Key Laboratory of Nuclear Physics and Technology, Peking University, Beijing 100871, China}
\affiliation{Center for High Energy Physics, Peking University, Beijing 100871, China}

\author{Haiyue Yu}
\affiliation{School of Physics and State Key Laboratory of Nuclear Physics and Technology, Peking University, Beijing 100871, China}

\author{Yanjie Zeng}
\affiliation{Institute of Theoretical
Physics, Chinese Academy of Sciences, Beijing 100190, China}
\affiliation{
School of Physical Sciences, University of Chinese Academy of Sciences, Beijing 100049, China}

\begin{abstract}

The Event Horizon Telescope (EHT) has significantly advanced our ability to study black holes, achieving unprecedented spatial resolution and revealing horizon-scale structures. Notably, these observations feature a distinctive dark shadow--primarily arising from faint jet emissions--surrounded by a bright photon ring. Anticipated upgrades of the EHT promise substantial improvements in dynamic range, enabling deeper exploration of low-background regions, particularly the inner shadow defined by the lensed equatorial horizon. Our analysis shows that observations of these regions transform supermassive black holes into powerful probes for annihilating dark matter, which is expected to accumulate densely in their vicinity. By analyzing the black hole image morphology and performing electron-positron propagation calculations in realistic plasma backgrounds derived from general relativistic magnetohydrodynamic simulations, we set stringent constraints on dark matter annihilation, requiring contributions below the astrophysical emission. These constraints, derived from both current EHT observations and projections for future upgraded arrays, exclude a substantial region of previously unexplored parameter space and remain robust against astrophysical uncertainties, including black hole spin and plasma temperature variations.

\end{abstract}

\date{\today}

\maketitle


{\it Introduction}--The rapid development of the very long baseline interferometry (VLBI) technique has enabled extraordinarily high angular resolution in radio astronomy. A notable illustration of this progress is the recent imaging of supermassive black holes (SMBHs) achieved by the Event Horizon Telescope (EHT)~\cite{EventHorizonTelescope:2019dse,EventHorizonTelescope:2022wkp}, which unveiled detailed astrophysical information in the strong gravity regions. Anticipated future upgrades of the EHT, including possible ground-based expansions and the proposed space-based Black Hole Explorer (BHEX), promise further enhancements in angular resolution, dynamic range, and baseline coverage~\cite{Johnson:2023ynn,Johnson:2024ttr}. These advancements are pivotal not only for astrophysical insights but also for exploring fundamental physics, including testing general relativity, examining black hole (BH) properties, and investigating new fundamental fields~\cite{Ayzenberg:2023hfw}.

One class of potential new particles, often considered as dark matter (DM) candidates, is found around the GeV mass scale. These candidates include weakly interacting massive particles (WIMPs)~\cite{Bertone:2010zza} and sub-GeV DM~\cite{Knapen:2017xzo}. In regions dominated by SMBHs' gravitational potential, the distribution of particlelike DM can steeply concentrate toward the BH, resulting in densities significantly higher than those near Earth~\cite{Gondolo:1999ef}.  Thus, SMBHs could serve as effective detectors for DM particles.

A promising approach for observing the DM spike is through indirect detection methods, like observing photons or cosmic rays resulting from potential DM annihilation~\cite{Gondolo:1999ef}. These methods leverage the principle that the annihilation rate, directly proportional to the square of the density, increases significantly in regions of high density, thereby enhancing the production of particles. At the EHT's millimeter radio band, photons are typically produced via synchrotron radiation, where electrons spiral around magnetic field lines~\cite{Falcke:1999pj}. Understanding the magnetic field structure near the horizon is essential for predicting radio fluxes from DM annihilation. In this study, we employ the best-fit general relativistic magnetohydrodynamic (GRMHD) simulation, which aligns with EHT's observations of SMBH M87$^*$~\cite{EventHorizonTelescope:2019ths,EventHorizonTelescope:2021bee}, specifically the magnetically arrested disk (MAD) model~\cite{Narayan:2003by}, to calculate the spectrum of DM annihilated electron-positron pairs near the SMBH and their resulting synchrotron radiation. Our results indicate that the density distribution of these pairs is similar in both equatorial and polar regions, contrasting with the MAD model prediction where electrons predominantly inhabit the disk, with lower densities in the jet region. Therefore, we utilize the morphology of BH images to impose constraints on DM annihilation, analyzing both current observations of the dark region resulting from faint jet emissions and future observations of the inner shadow--a region delineated by the lensed image of the equatorial horizon~\cite{Chael:2021rjo}. These constraints, based on the morphology, prove significantly more stringent than those derived from the total intensity of the image~\cite{Lacroix:2016qpq,Yuan:2021mzi}.


{\it Electron-Positron Spectrum from Dark Matter Annihilations Near Supermassive Black Holes}--The density of DM within a galaxy is typically highest at its center, and the presence of an SMBH can further sharpen this distribution, leading to an increased density toward the center. When considering an initial DM distribution profile where the mass of the central SMBH is sufficiently low to influence the distribution, the slow accretion of ordinary matter onto the SMBH can lead to an adiabatic distortion of DM phase space by the SMBH's gravitational potential~\cite{Gondolo:1999ef}. For instance, in a standard Navarro-Frenk-White (NFW) profile derived from N-body simulations, which neglect the SMBH, the energy density $\rho(r) \propto r^{-1}$ in the central region, where $r$ represents the distance from the galaxy's center~\cite{Navarro:1995iw}. The presence of an SMBH, assuming its adiabatic growth, leads to the formation of a DM spike, with its density scaling as $\propto r^{-7/3}$ at the center and transitioning back to the NFW profile in the outer regions. However, this adiabatic formation of a spike can be mitigated by factors such as stellar interactions~\cite{Vasiliev:2008uz}. Furthermore, DM annihilation introduces an upper limit on the density, inversely proportional to $\langle\sigma v\rangle$, the thermally averaged cross section times the relative velocity of DM particles, which effectively regulates the central spike into a flat core~\cite{Gondolo:1999ef}.

Depending on their mass and interaction channels with standard model particles, DM can annihilate into a spectrum of final-state particles, notably electrons and positrons in this study, which emit synchrotron radiation in magnetic fields~\cite{Cirelli:2010xx}. These particles, after being produced near an SMBH, move under the influence of both gravitational and magnetic fields. We have developed a comprehensive framework, detailed in Supplemental Material~\cite{SupplementalMaterial}, to calculate the steady-state electron-positron spectrum after propagation in the complex magnetic field environment surrounding an SMBH. We also include a discussion of the approximations employed, such as the omission of certain general relativistic effects, which have only a minor impact on our results. This framework represents an advancement over prior spherical models of DM propagation~\cite{Regis:2008ij,Lacroix:2013qka,Lacroix:2016qpq,Yuan:2021mzi}, aiming for a more astrophysically precise description.

\begin{figure}[t]
    \centering
    \includegraphics[width=0.48\textwidth]{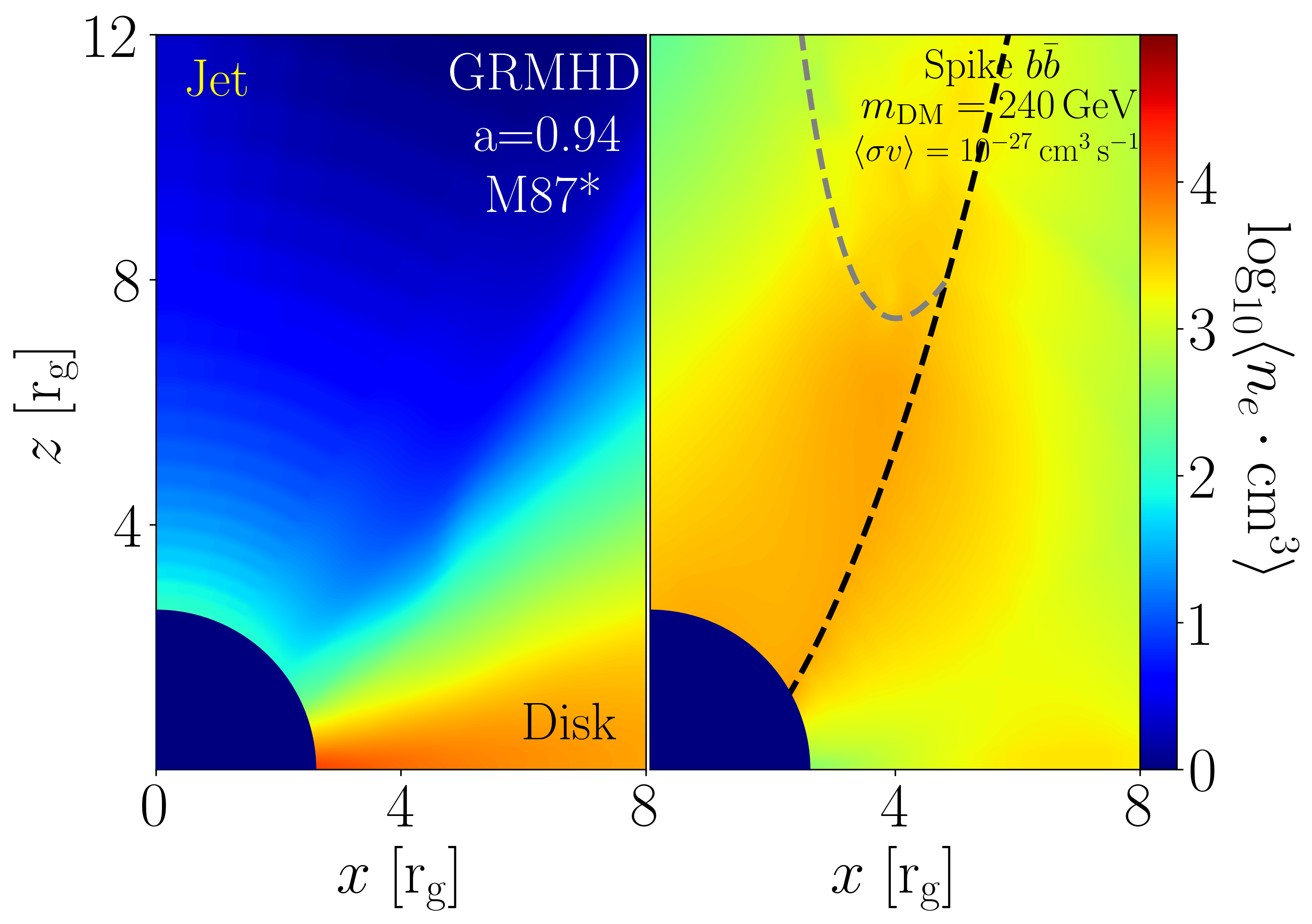}
    \caption{Electron-positron number density $n_e$ from an azimuthally averaged MAD model based on GRMHD (left) versus from DM annihilation (right). The $z$ axis is aligned with the BH spin, adopting a BH spin of $a_J = 0.9375$ in the GRMHD simulation. The DM profile assumes a standard spike originating from an initial NFW profile, with a DM mass of $m_{\rm DM}\!=\! 240\,\mathrm{GeV}$, annihilation via the $b\bar{b}$ channel, and a cross section of $\langle \sigma v \rangle = 10^{-27}\,\mathrm{cm}^3\,\mathrm{s}^{-1}$. The black dashed line marks the jet boundary, while the gray dashed line indicates the stagnation surface.}
\label{Fig:1}
\end{figure}

The background magnetic field utilized for calculating the propagation of electrons and positrons is derived from MAD-type models~\cite{Narayan:2003by}, which is supported by observations of the SMBH M87$^*$ by the EHT~\cite{EventHorizonTelescope:2019ths,EventHorizonTelescope:2021bee}. In Fig.~\ref{Fig:1}, we show an example of the azimuthally and time-averaged electron and positron density distribution, $n_e$, for a BH spin of $a_J = 0.9375$. The $z$ axis is aligned with the BH’s spin. The density is primarily concentrated in the disk region, with significantly lower values in the jet region. We utilize an axisymmetric ansatz to fit the magnetic field configurations in both the disk and jet regions, based on GRMHD simulations for the MAD. We assume that the bulk velocities of the electron and positron plasma resulting from DM annihilation align with the magnetic field lines. In the disk region, their radial components are directed toward the BH, with a magnitude scaling as $(2\,r_g/r)^{1/2}$, where $r_g$ is the gravitational radius of the BH. In the jet region, we account for the rotation of the magnetic field by incorporating both the inertial centrifugal potential in the corotating frame and the gravitational potential, leading to the identification of a zero-velocity surface known as the stagnation surface~\cite{Pu:2015rja,Pu:2017akw,Pu:2020pky}. The initial spectra of DM annihilated electron and positron pairs are calculated using the PPPC~\cite{Cirelli:2010xx} and MadDM~\cite{Ambrogi:2018jqj} packages.

The right panel of Fig.~\ref{Fig:1} presents an example of the resulting electron and positron density, excluding electrons from the GRMHD MAD model, where the DM mass is $m_{\rm DM}\!=\!240\,$GeV, with the annihilation channel being $b\bar{b}$ and $\langle\sigma v\rangle = 10^{-27}\,\text{cm}^3\text{s}^{-1}$. The black dashed line marks the jet boundary, while the gray dashed line indicates the stagnation surface. The densities in the jet and disk regions prove to be comparable, differing markedly from the GRMHD MAD case. This discrepancy arises because the distribution of electrons and positrons is expelled by the jet in the GRMHD scenario, whereas DM annihilation continuously supplies them within the jet cone.


\begin{figure*}[t]
    \centering
   \includegraphics[width=0.9\textwidth]{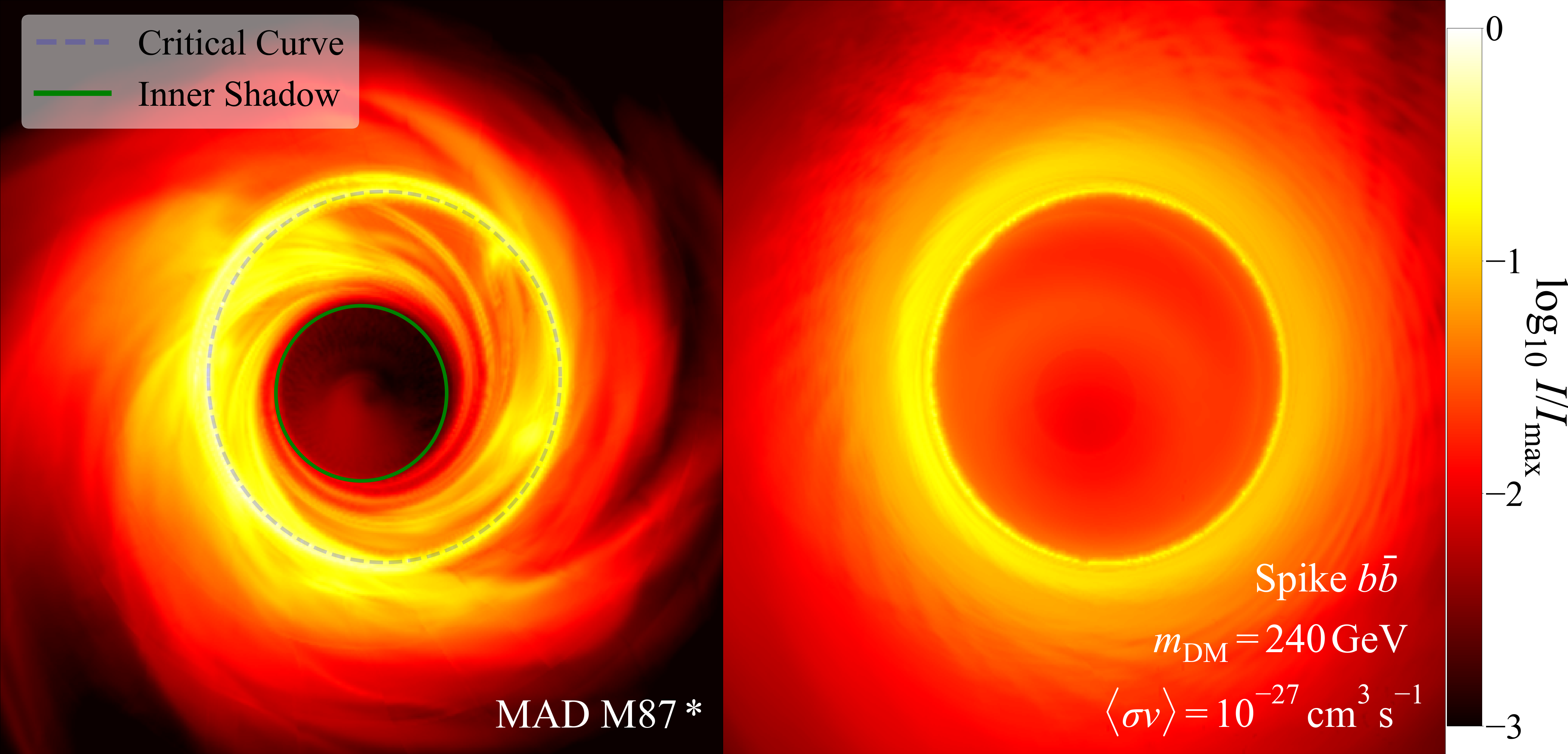} 
   \includegraphics[width=0.908\textwidth]{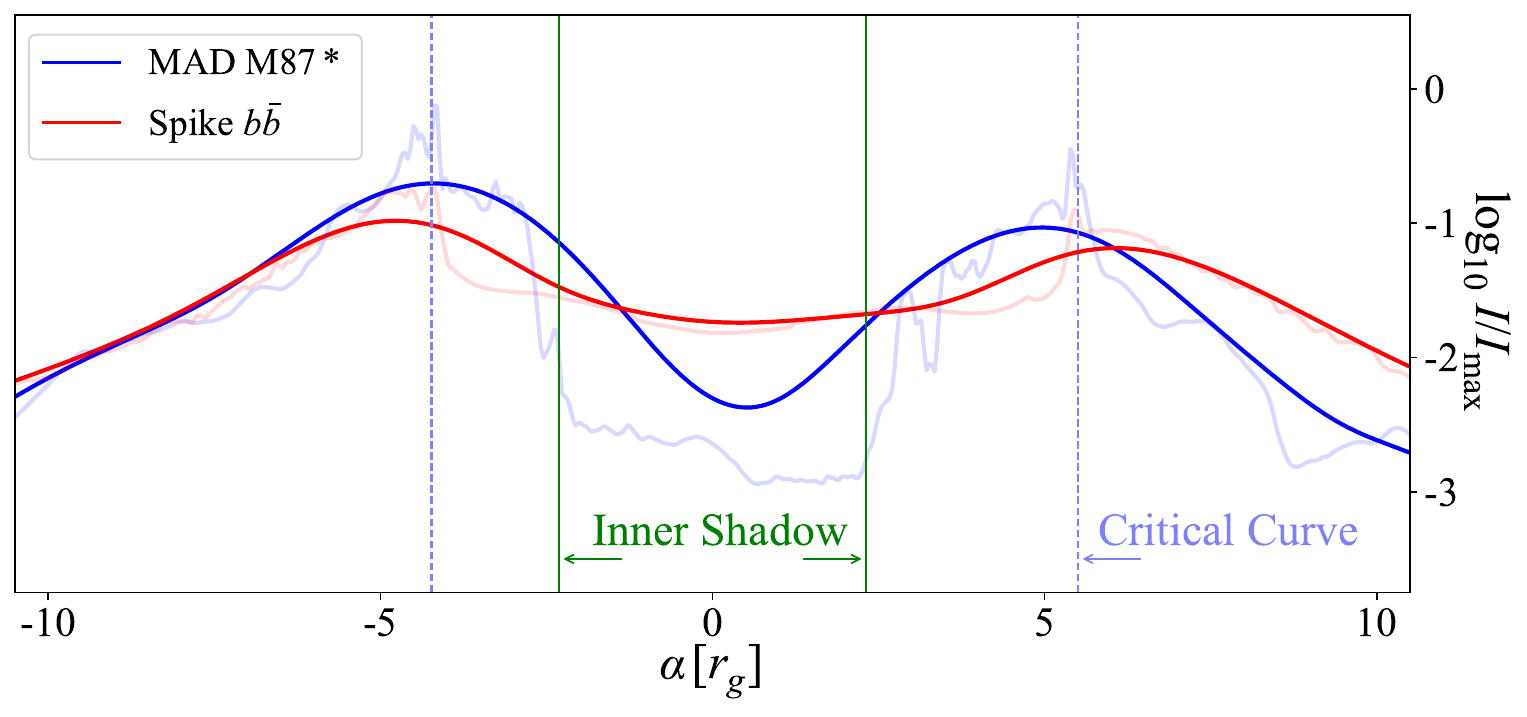}\ \
  \caption{Top: normalized intensity map ($I/I_{\rm max}$) in logarithmic scale comparing the MAD model from GRMHD (left) to DM annihilation (right), for the SMBH M87$^*$ with $a_J = 0.9375$ at an inclination angle of $\theta_o = 163^\circ$. Normalization is against the maximum intensity in the MAD model, $I_{\rm max}$. The critical curve and the inner shadow's boundary are denoted by dashed and green lines, respectively. Both maps exhibit a total intensity of $0.6$\,Jy.
    Bottom: one-dimensional normalized intensity profile along the axis ($\alpha$), centered on the BH and perpendicular to the spin projection, showcasing contrasts between GRMHD MAD (blue) and DM annihilation (red) scenarios. Dark and light shades indicate the original image and its Gaussian-smoothed counterpart with an $1\,r_g$ kernel, respectively.}
    \label{Fig:2}
\end{figure*}

{\it Constraints from the EHT and future EHT upgrades}--Constraints on DM annihilation require that radiation from this process remains below the astrophysical background, positioning astrophysical radiation as a pivotal factor in determining the exclusion parameter space. We focus on the synchrotron radiation from the MAD model at $230$\,GHz radio frequency band. Our analysis concentrates on the synchrotron radiation emanating from the MAD model within the $230$\,GHz radio frequency band. By employing the \texttt{RAPTOR} covariant radiative transfer package~\cite{Bronzwaer:2018lde}, we generate a horizon-scale intensity ($I$) map for MAD outside an SMBH with an inclination angle of $\theta_o = 163^\circ$, consistent with observations of M87$^*$, depicted in the top left panel of Fig.~\ref{Fig:2}. Subsequent application of Gaussian smearing with an approximately $2\,r_g$ kernel to simulate the current EHT's angular resolution yields an image that closely matches the observed data~\cite{EventHorizonTelescope:2019ths}, with a total flux of approximately $0.6$\,Jy.

The BH image, presented on a logarithmic scale, reveals fascinating morphological features that reflect the interplay between Kerr spacetime and the sources of emission. The dashed line in Fig.~\ref{Fig:2} delineates the critical curve, beyond which null geodesics are captured into bound orbits around the BH, creating a photon ring from light circling multiple times~\cite{1965SvA.....8..868P,1968ApJ...151..659A,Luminet:1979nyg}. Consequently, the intensity observed in the adjacent region results from the cumulative effect of photons undergoing multiple loops, achieving their peak in the map~\cite{Johannsen:2010ru}. Within the critical curve, backward geodesics terminate at the BH horizon, known as the BH ``shadow''~\cite{Falcke:1999pj}. Nonetheless, this area is not completely devoid of light, as emission, proportional to electron density $n_e$ and magnetic field strength $B$, reaches up to the horizon. As shown in Fig.~\ref{Fig:1} and Supplemental Material, the distribution of $B$ slightly changes with the polar angle, whereas $n_e$ is predominantly found in the accretion disk, with much lower concentrations in the jet region. This distribution creates a geometrically thin emission profile mainly in the equatorial disk, leading to sharp intensity depression in the central area, identified as the ``inner shadow''~\cite{Chael:2021rjo}. The boundary of this inner shadow, marked in green in Fig.~\ref{Fig:2}, outlines the lensed contour of the BH's equatorial horizon.

The current EHT's capability to detect the inner shadow is hindered by its limited dynamic range ($\sim 10$) and angular resolution on the intensity dip. In contrast, future EHT upgrades are anticipated to achieve a substantial enhancement in both dynamic range ($\sim 1000$) and angular resolution, equivalent to a Gaussian kernel of approximately $1\,r_g$. This improvement holds the promise of capturing this fascinating feature and delving into the BH horizon~\cite{Chael:2021rjo}. The bottom panel of Fig.~\ref{Fig:2} displays the one-dimensional intensity distribution along an axis that crosses the BH center and is perpendicular to the projection of its spin. The use of dark and light shades serves to distinguish between the original image and the image after applying Gaussian smearing. Notably, within the inner shadow, the foreground emission from the jet region is significantly suppressed--by roughly 2 orders of magnitude compared to the peak intensity--and, thus, well within the detection capabilities of the proposed EHT upgrades~\cite{Chael:2021rjo}.

The pronounced dimming within the shadow region establishes a solid framework for testing DM annihilation. As shown in Fig.~\ref{Fig:1}, the electron-positron density $n_e$, induced by DM annihilation, attains comparable levels in both the jet and disk regions, akin to the results observed in spherical emission models~\cite{Vincent:2022fwj}. This morphology allows the DM-induced density to exceed the jet's emission in the shadow area. The intensity map, derived solely from DM annihilation and presented in the top right panel of Fig.~\ref{Fig:2} with the benchmark parameters from Fig.~\ref{Fig:1}, reveals a total intensity nearly matching the astrophysical emission of $0.6\,$Jy. In contrast, the inner shadow is significantly illuminated, lacking the expected intensity reduction. This discrepancy becomes even more evident in the bottom panel of Fig.~\ref{Fig:2}, where the red line for DM annihilation markedly exceeds the expected intensity levels within the inner shadow, deviating from the astrophysical emission patterns.

\begin{figure}[t]
    \centering
  \includegraphics[width=0.48\textwidth]{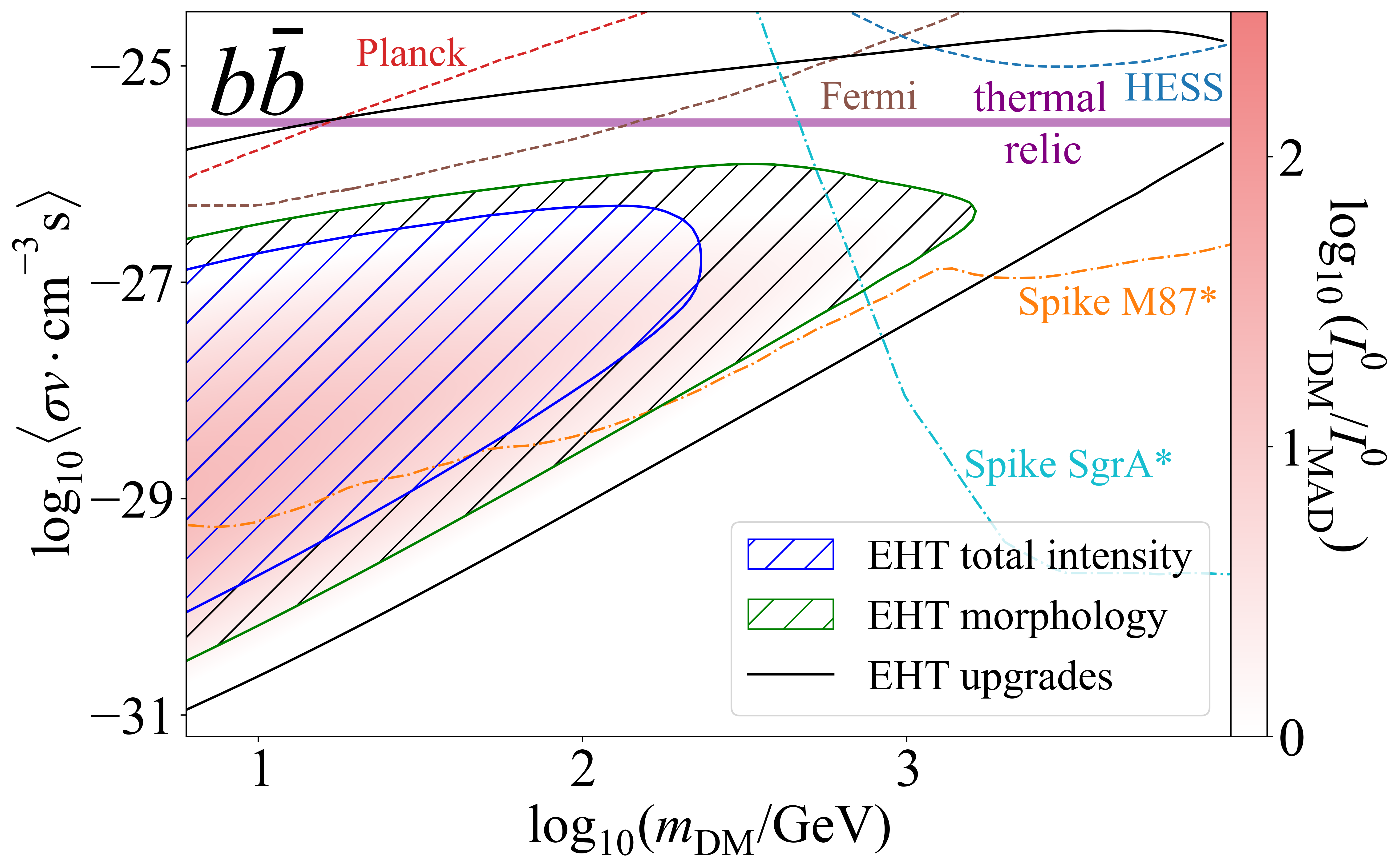}
   \includegraphics[width=0.48\textwidth]{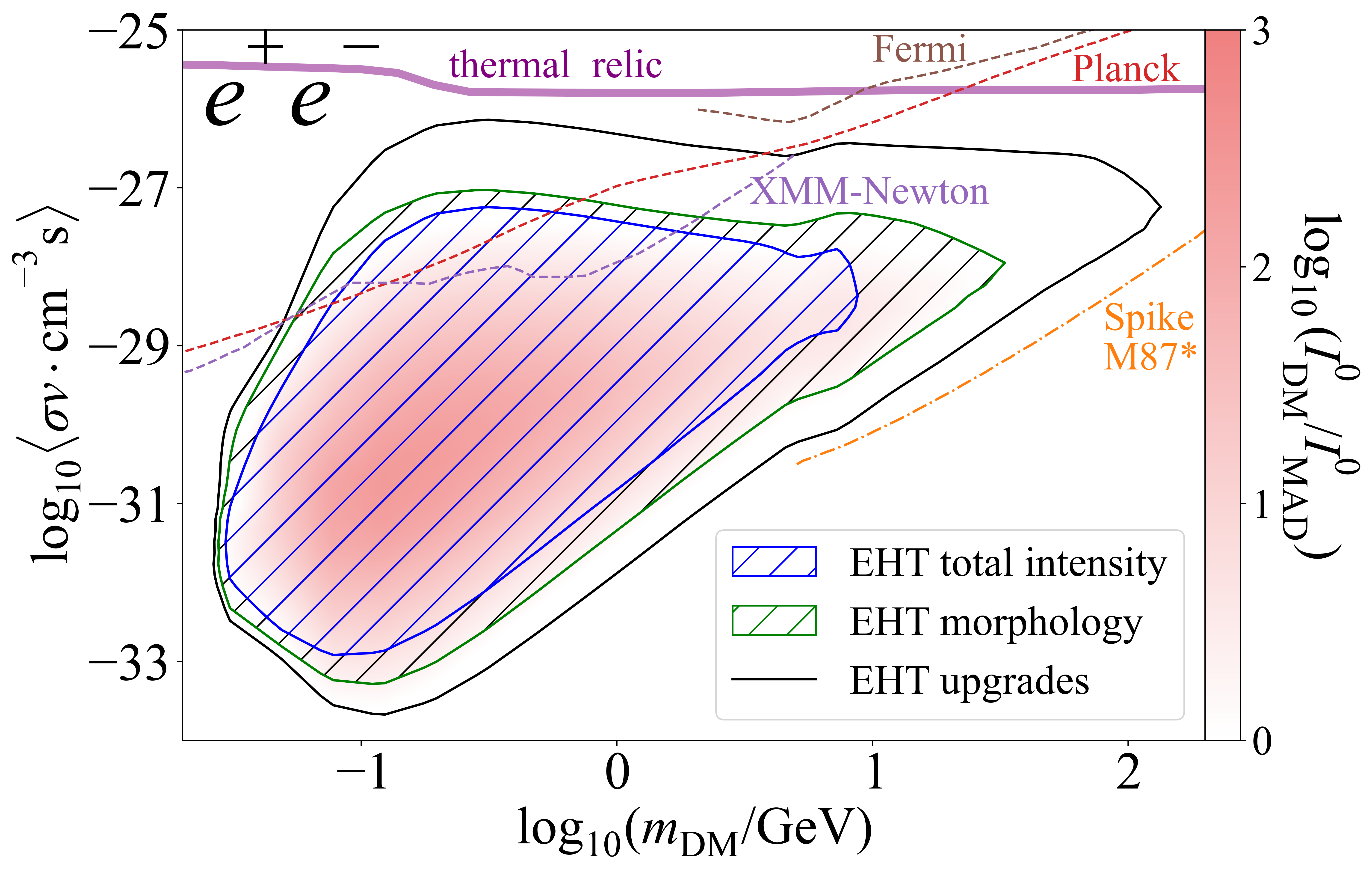}
\caption{Constraints on DM annihilation channels $b\bar{b}$ (top) and $e^+e^-$ (bottom), derived from current EHT observations (diagonal stripes), which include criteria based on total intensity (blue) and BH image morphology (green), as well as projections for the EHT upgrades (black). The pink shades illustrate the ratio of local intensity between DM annihilation and astrophysical emissions from the MAD model in the inner shadow. The thermal relic annihilation cross section~\cite{Steigman:2012nb} is depicted by the purple line. Previous constraints from various sources, including Planck CMB observations~\cite{Slatyer:2015jla}, Fermi-LAT~\cite{Fermi-LAT:2015att}, H.E.S.S.~\cite{HESS:2016mib}, AMS~\cite{Bergstrom:2013jra}, and XMM-Newton~\cite{Cirelli:2023tnx,DelaTorreLuque:2023olp}, as well as spike-based constraints~\cite{Lacroix:2015lxa,Balaji:2023hmy}, are shown for comparative analysis in dashed and dot-dashed lines, respectively.}
\label{Fig:3}
\end{figure}

In Fig.~\ref{Fig:3}, we illustrate the exclusion regions derived from analyses using both the current EHT observations and the future prospects, targeting DM annihilation channels $b\bar{b}$ and $e^+e^-$, associated with a DM spike around M87$^*$. The regions with diagonal stripes represent constraints from current EHT observations. The regions enclosed by the blue lines represent exclusion constraints, ensuring that the total intensity from DM annihilation does not exceed the overall astrophysical emission within the EHT field of view, which extends to approximately $16\,r_g$ from the BH. The morphological constraints, shown within the green lines, impose a stricter criterion by demanding that the local intensity from DM annihilation, after Gaussian smearing with a $2\,r_g$ kernel, consistently remains below the astrophysical emission within $10\,r_g$ and exceeds the dynamic range threshold of $1/10$. The prospective morphological constraints from the EHT upgrades, depicted in black, significantly expand beyond that established by the current EHT, underscoring the increased sensitivity to the BH image’s morphology. The density within the exclusion zone represents the ratio of local intensity between DM annihilation and MAD-induced emissions in the inner shadow region. The EHT upgrades' exclusion region primarily arises from the enhanced dynamic range of approximately $1000$. In the $b\bar{b}$ channel, the exclusion region closely aligns with areas exhibiting the highest electron-positron densities as shown in Supplemental Material. In contrast, the exclusion contour for the $e^+e^-$ channel exhibits a notable turnover near $0.1$\,GeV, due to the reduced efficiency of synchrotron radiation at $230$\,GHz for $m_{\rm DM}$ below this threshold.

Several astrophysical uncertainties can influence our results, even within the MAD model class favored by multiple EHT observations of M87$^*$~\cite{EventHorizonTelescope:2019pgp,EventHorizonTelescope:2021srq,EventHorizonTelescope:2023gtd}. One such parameter is the BH spin $a_J$, which is currently favored to be high~\cite{EventHorizonTelescope:2019pgp}, but has only a lower bound of $a_J \geq 0.5$~\cite{Cruz-Osorio:2021cob}. While the main text presents results using $a_J = 0.9375$, we also perform a full analysis for $a_J = 0.5$. The resulting constraints, shown in Supplemental Material, vary only slightly from the fiducial case, demonstrating the robustness of our conclusions. We additionally explore variations in plasma-related parameters such as $R_{\rm high}$ and $R_{\rm low}$, which characterize the proton-to-electron temperature ratio in regions of weak and strong magnetization, respectively~\cite{Moscibrodzka:2015pda,Bronzwaer:2018lde}. These variations likewise show minimal impact on our exclusion limits.

The purple line in Fig.~\ref{Fig:3} denotes the theoretically well-motivated thermal relic annihilation cross section~\cite{Steigman:2012nb}, which the EHT upgrades could potentially probe for DM masses up to approximately $10$\,TeV. For context, the figure also presents previous constraints obtained from the Planck observation of the cosmic microwave background (CMB)~\cite{Slatyer:2015jla}, as well as constraints from the Fermi Large Area Telescope (Fermi-LAT)~\cite{Fermi-LAT:2015att}, the High Energy Stereoscopic System (H.E.S.S.)~\cite{HESS:2016mib}, the Alpha Magnetic Spectrometer Experiment (AMS)~\cite{Bergstrom:2013jra}, and the X-ray Multi-Mirror Mission (XMM-Newton)~\cite{Cirelli:2023tnx,DelaTorreLuque:2023olp}, depicted with dashed lines. Additionally, dot-dashed lines highlight previous constraints that are specific to a DM spike profile~\cite{Lacroix:2015lxa,Balaji:2023hmy}. Notably, some previous constraints derived from synchrotron radiation may lead to an overestimation of the exclusion region, attributed to the adoption of a simplified magnetic field profile~\cite{Lacroix:2015lxa}. Our constraints improve upon these existing bounds primarily due to the EHT’s superior angular resolution and dynamic range, which enable probing DM annihilation in a compact region with minimal astrophysical foreground contamination.


{\it Discussion}--The exceptional angular resolution offered by the EHT and its future upgrades not only enriches our understanding of astrophysical phenomena but also provides robust constraints on fundamental particle physics~\cite{Chen:2019fsq,Yuan:2020xui,Chen:2021lvo,Chen:2022nbb,Chen:2022kzv,Ayzenberg:2023hfw}. This study illustrates how the detailed morphology of BH images, particularly the shadow regions, can inform constraints on DM annihilation. We capitalize on the heightened DM density near SMBHs and the unique characteristics of BH images--most notably, the diminished intensity in the inner shadow area, a result of the jet's sparse electron density. To ensure the reliability of our constraints, we have developed an advanced framework for simulating the propagation of electrons and positrons within a background fit derived from a realistic GRMHD profile. This profile is consistent with the latest intensity and polarimetric observations and represents an advancement beyond previously assumed simpler spherical accretion models~\cite{Regis:2008ij,Lacroix:2013qka,Lacroix:2016qpq,Yuan:2021mzi}. While our analysis uses M87$^*$ as a primary example, our discussions and constraints are equally applicable to another EHT focus, Sgr A$^*$, within the Milky Way--given that observations similarly support a MAD profile around it and an almost direct line of sight~\cite{EventHorizonTelescope:2022wkp,GRAVITY:2023avo}. Note that the DM considered here contributes solely to intensity enhancement, with negligible effects on geodesics due to its minimal mass distribution near the horizon. This distinguishes it from other dark sector phenomena, such as superradiant clouds~\cite{Sengo:2022jif,Chen:2022kzv}, which can reach up to $10\%$ of the BH’s mass, or BH alternatives~\cite{Sengo:2024pwk,Huang:2024bbs}.

Looking ahead, this research aims to extensively explore DM characteristics, including an expanded range of annihilation channels, a wider spectrum of mass scales, and exploring phenomena such as $p$-wave annihilation~\cite{Christy:2023tdv} and forbidden DM~\cite{Cheng:2022esn}, both of which are expected to be significantly amplified in the vicinity of SMBHs. Furthermore, our analysis of the horizon-scale intensity map could be broadened to assess morphological changes at larger scales, encompassing extended jet structures and the accretion flow at greater distances from the BH. In these areas, the electron density from DM annihilation with a large cross section decreases more slowly than the astrophysical background.

From an observational standpoint, the incorporation of linear and circular polarization intensity measurements could unveil new perspectives in differentiating between electrons and positrons~\cite{Anantua:2019bna,Emami:2021ick,Emami:2022ycp,Anantua:2023mjk}. This differentiation is crucial since astrophysical plasmas typically exhibit a significant deficit in positron populations, potentially facilitating the imposition of tighter constraints. The advent of forthcoming multifrequency observations~\cite{Chael:2022meh} is expected to further improve detection capabilities.


{\it Acknowledgments}--We are grateful to Richard Anantua, Marco Cirelli, Razieh Emami, Jordan Koechler, George Wong, Xiao Xue, Daixu Yang and Yue Zhao for useful discussions. 
The Center of Gravity is a Center of Excellence funded by the Danish National Research Foundation under Grant No. 184.
Y. C. acknowledges support by VILLUM Foundation (Grant No. VIL37766) and the DNRF Chair program (Grant No. DNRF162) by the Danish National Research Foundation, the European Union’s H2020 ERC Advanced Grant “Black holes: gravitational engines of discovery” Grant Agreement No. Gravitas–101052587, the Rosenfeld foundation in the form of an Exchange Travel Grant, and the COST Action COSMIC WISPers CA21106, supported by COST (European Cooperation in Science and Technology). This project has received funding from the European Union's Horizon 2020 research and innovation programm under the Marie Sklodowska-Curie Grant Agreement No. 101007855 and Grant Agreement No. 101131233. 
R.D. is supported in part by the National
Key R\&D Program of China (No. 2021YFC2203100).
Y.M. is supported by the National Natural Science Foundation of China (Grant No. 12273022), the Shanghai Municipality orientation program of Basic Research for International Scientists (Grant No. 22JC1410600), and the National Key R\&D Program of China (No. 2023YFE0101200).
J.S. is supported by the National Key Research and Development Program of China under Grants No. 2020YFC2201501 and No. 2021YFC2203004, Peking University under startup Grant No. 7101302974, and the NSFC under Grants No. 12025507, No. 12150015, and No. 12450006.

Views and opinions expressed are however those of the author only and do not necessarily reflect those of the European Union or the European Research Council. Neither the European Union nor the granting authority can be held responsible for them.

%

\pagebreak
\widetext
\begin{center}
\textbf{\large Supplemental Materials: Illuminating Black Hole Shadow with Dark Matter Annihilation}
\end{center}
\setcounter{equation}{0}
\setcounter{figure}{0}
\setcounter{table}{0}
\setcounter{section}{0}
\makeatletter
\renewcommand{\theequation}{S\arabic{equation}}
\renewcommand{\thefigure}{S\arabic{figure}}
\renewcommand{\bibnumfmt}[1]{[#1]}
\renewcommand{\citenumfont}[1]{#1}

\hspace{5mm}

The Supplemental Material provides a comprehensive procedure for calculating the electron-positron spectrum outside the black hole (BH) resulting from dark matter (DM) annihilation, as well as the intensity map derived from synchrotron radiation emitted by these electrons and positrons. Additionally, the formalism for the critical curve and the contour of the inner shadow is included. We also demonstrate that astrophysical uncertainties have only a minor impact on our results.
Throughout this study, we employ natural units.

\section{Electron-Positron Spectrum from Dark Matter Annihilations near Supermassive Black Holes}
This section examines the propagation of charged particles emerging from DM pair annihilations near supermassive BHs (SMBHs). We emphasize the significant role of electrons and positrons as primary contributors to synchrotron emission. Their propagation is analyzed in terms of energy gains from adiabatic accretion influenced by the SMBH's gravitational potential, alongside energy losses through various radiative mechanisms, culminating in a steady-state distribution in phase space.

\subsection{Propagation of Electrons and Positrons}

After being generated from DM annihilations, the flow of electrons and positrons navigates through an environment shaped by astrophysical accretion flows and the gravitational potential of the SMBH. The steady-state phase space distribution of these electrons and positrons from DM annihilation, denoted as $f_e(\vec{r},\,\vec{p}\,)$, is defined at position $\vec{r}$ with momentum $\vec{p}$. This distribution adheres to the transport equation~\cite{Aloisio:2004hy,Regis:2008ij}:
\begin{align}
\nabla \cdot (\vec{v}_b\,f_e) + \nabla_p \cdot (\dot{\vec{p}}_{\rm adi}\,f_e) + \nabla_p \cdot (\dot{\vec{p}}_{\rm rad}\,f_e)+\nabla \cdot (D_{xx} \nabla f_e)+ \nabla_p \cdot (D_{pp} \nabla_p f_e)=q(\vec{r},p).
\label{eq:transport_eq}
\end{align}
Here, $\vec{v}_b$ represents the bulk flow velocity of the plasma, while $\dot{p}_{\rm adi}$ and $\dot{p}_{\rm rad}$ denote the momentum change rates due to adiabatic compression/expansion and radiation, respectively. The terms $D_{xx}$ and $D_{pp}$ correspond to the spatial and momentum diffusion coefficients, and $q(\vec{r},p)$ is the source term for DM annihilation. The first two terms describe advective transport and adiabatic compression, while the remaining terms account for radiative losses and diffusion.

We can simplify Eq.~(\ref{eq:transport_eq}) using the isotropic momentum approximation, as the relativistic electrons and positrons far exceed the bulk flow velocity $\vec{v}_b$, and both the DM source function $q(\vec{r},p)$ and the relevant momentum change processes are isotropic in momentum space.

Below we analyze each component systematically. Readers primarily interested in the final simplified equation may proceed to Eq.~(\ref{eq:expanded}).
 \begin{itemize}
\item \textbf{Adiabatic Compression} The momentum gain from plasma compression is:
\be
        \dot{p}_{\rm adi} = -p(\nabla\cdot\vec{v}_b\,)/3 \approx 1.03\times 10^{-6} \,{\rm GeV\,s^{-1}} \left(\frac{p}{1 \mathrm{GeV}}\right) \left( \frac{r}{r_g} \right)^{-1} \left( \frac{v_b}{0.1}\right)\,,
\ee
where $r_g = 3.1 \times 10^{-4}$\,pc is the gravitational radius of M87$^*$, and we approximate $-\nabla \cdot \vec{v}_b \approx v_b/r$.

\item \textbf{Radiative Losses}
The total radiative loss encompasses four processes:
$\dot{p}_{\rm rad}=\dot{p}_{\rm syn}+\dot{p}_{\rm brem}+\dot{p}_{\rm IC}+\dot{p}_{\rm C}$, which include synchrotron radiation ($\dot{p}_{\rm syn}$), bremsstrahlung ($\dot{p}_{\rm brem}$), inverse Compton scattering ($\dot{p}_{\rm IC}$), and Coulomb collisions ($\dot{p}_{\rm C}$). Their approximations are given as~\cite{Regis:2008ij,Chael:2017ahn}:
\be \begin{split}
\dot{p}_{\rm rad} &=\dot{p}_{\rm syn}+\dot{p}_{\rm brem}+\dot{p}_{\rm IC}+\dot{p}_{\rm C},\\
    \dot{p}_{\rm syn}&\approx -2.01\times 10^{-7} \,{\rm GeV\,s^{-1}} \left(\frac{B}{1\,\rm G}\right)^2\left(\frac{p}{1\,\rm GeV}\right)^2 \,,\\
    \dot{p}_{\rm brem}&\approx -1.37\times 10^{-16}\,{\rm GeV\,s^{-1}}\left(\frac{\bar{n}_{ion}}{1\,\rm cm^{-3}}\right)\left(\frac{p}{1\,\rm GeV}\right)\left(\ln\frac{p}{1\,\rm GeV}+7.94\right)\,,\\
    \dot{p}_{\rm IC}&\approx -1.02\times 10^{-16}\,{\rm GeV\,s^{-1}}\left(\frac{\bar{u}_{\rm r}}{1\,\rm eV\,\rm cm^{-3}}\right)\left(\frac{p}{1\,\rm GeV}\right)^2\,,\\
    \dot{p}_{\rm C}&\approx -7.62\times10^{-18}\,{\rm GeV\,s^{-1}}\left(\frac{\bar{n}_{e}}{1\,\rm cm^{-3}}\right)\left[\ln\frac{p}{1\,\rm GeV}+\ln\frac{\bar{n}_{e}}{1\,\rm cm^{-3}}+82.3\right]\,,
\end{split} \label{eq:pdot}\ee
Here, we consider typical astrophysical conditions, including a magnetic field strength of $B \sim 1\,\rm G$, a background ion and thermal electron number density of $\bar{n}_{ion/e}\sim 1\,\rm cm^{-3}$, and a background radiation energy density of $\bar{u}_{\rm r}\sim 1\,\rm eV\,cm^{-3}$. Under these conditions, the synchrotron energy loss dominates, allowing us to neglect the latter three effects.

\item \textbf{Spatial Diffusion}
Unlike the plasmas in GRMHD simulations, where the dynamics are dominated by much heavier ions with larger Larmor radii, the lighter electron-positron plasmas considered here are far less susceptible to diffusion. The spatial diffusion scale, defined as the distance over which charged particles lose most of their energy while diffusing, is given by~\cite{Regis:2008ij}:
\begin{align}
d_L \equiv \sqrt{\frac{D_{xx} p}{\dot{p}}}
\approx 4.18 \times 10^{-4} \, r_g \, \left(\frac{B}{1\mathrm{G}}\right)^{-3/2} \ll r_g
\end{align}
Here, the spatial diffusion coefficient follows from Bohm diffusion as $D_{xx} = p/(3 e B)$, where energy losses are primarily due to synchrotron radiation. Compared to synchrotron losses, the effect of spatial diffusion is suppressed by $d_L/r_g$ and can thus be neglected.

\item \textbf{Momentum Diffusion} Momentum diffusion, caused by magnetic turbulence, leads to stochastic acceleration by Alfvén waves, contributing to momentum gain. This process is expressed as~\cite{Dermer:1995ju}:
\begin{align}
\dot{p}_{\rm acc}=\frac{m_e}{p^2}\frac{\partial}{\partial p}\left(p^2 D_{pp}\right)\sim m_e\beta_A^2 \zeta \,r_L^{\alpha-2}\lambda_{\rm max}^{1-\alpha}p^{\alpha-1} \approx 1.52 \times 10^{-9}\,{\rm GeV\,s^{-1}} \beta_A^2 \left( \frac{\zeta}{0.1}\right) \left( \frac{10^4 \, r_g}{\lambda_{\rm max}}\right) \left( \frac{p}{\mathrm{GeV}}\right)\,.
\end{align}
Here, the momentum diffusion coefficient is given by $D_{pp} \sim \beta_A^2 \zeta r_L^{\alpha-2} \lambda_{\rm max}^{1-\alpha} p^{\alpha}$, where the magnetic turbulence follows a power-law spectrum with index $\alpha = 2$. The parameter $\beta_A \approx 1$ represents the Alfv\'en  velocity, $\zeta \equiv \delta B^2/B^2 \sim 0.1$ quantifies the ratio of turbulent to average magnetic field strength, $r_L \equiv m_e/(eB)$ is the electron Larmor radius, and $\lambda_{\rm max} \sim 10^4 r_g$ denotes the maximum wavelength of Alfv\'en waves. This contribution remains small compared to $\dot{p}_{\rm adi}$ and $\dot{p}_{\rm syn}$ and can therefore be neglected.
\end{itemize}

Synthesizing the discussion above, we retain only the first three terms in Eq.~(\ref{eq:transport_eq}) and adopt the isotropic momentum approximation, simplifying Eq.~(\ref{eq:transport_eq}) to
\begin{align}
    \vec{v}_b \cdot \nabla  f_e+(\dot{p}_{\rm adi}+\dot{p}_{\rm syn})\frac{\partial f_e}{\partial p}=q(\vec{r},p)-\frac{4 \, \dot{p}_{\rm syn} \, f_e}{p},
    \label{eq:expanded}
\end{align}
The evolution of the electron-positron distribution follows specific streamlines determined by the bulk velocity $\vec{v}_b$. Due to their high charge-to-mass ratios relative to heavier ions in the accretion flow, electrons and positrons closely follow magnetic field lines. Consequently, the streamline direction $\hat{v}_b(\vec{r}\,)$ aligns with the magnetic field direction, denoted as $\hat{B}(\vec{r}\,)$. The bulk velocity profiles considered in this study will be discussed in detail in Secs.~\ref{sec:BfGRMHD} and \ref{sec:BVF}.

Next, we consider the solution to Eq.~(\ref{eq:expanded}). The momentum evolution along a streamline can be parameterized as
\begin{align}
\frac{\d p}{\d s}=\frac{\dot{p}_{\rm adi}+\dot{p}_{\rm syn}}{v_b},
\label{eq:dp_ds}
\end{align}
which can be directly integrated to obtain
\begin{align}
p-p_{i}(s_{i}) = \int_{s_{i}}^{s_{\vec{r}}} \d s \, \frac{\dot{p}_{\rm adi}+\dot{p}_{\rm syn}}{v_b}.
\label{eq:pinjint}
\end{align}
Here, $s$ represents the distance traveled along the streamline, starting from an injection point $s_i$ with initial momentum $p_i$ due to DM annihilation, to a final point $s_{\vec{r}}$ at location $\vec{r}$. For a given momentum $p$ at $\vec{r}$, the corresponding injection momentum $p_i$ can be determined along the streamline.

The left-hand side of Eq.~(\ref{eq:expanded}) can be rewritten as $\d f_e/\d s$ along the streamline. The solution is then obtained by summing the contributions from all upstream injection points $s_i$ propagating to $\vec{r}$:
\be\begin{split}
    f_e(\vec{r},p) &=\int^{s_{\vec{r}}}_{s_{b}}  \d s_{i}\, \frac{q\,(s_{i},p_{i}(s_{i}))}{v_b(s_{i})} \exp \left[ \int^{s_{i}}_{s_{\vec{r}}}  \d {s} \, \frac{4 \,\dot{p}_{\rm syn}(s,p_{i}({s}))}{p_{i}({s}) \, v_b({s})} \right]\,\\
    &= \int^{s_{\vec{r}}}_{s_{b}}  \d s_{i} \frac{q\,(s_{i},p_{i}(s_{i}) )}{v_b(s_{i})} \left(\frac{p_{i}(s_{i})}{p}\right)^4 G(s_{\vec{r}})\, G^{-1}(s_{i}).
    \label{eq:f_solution}
\end{split}\ee
Here, $s_b$ denotes the most distant injection point along the streamline, serving as a boundary condition. The function $G$ is a form factor that satisfies
\begin{align}
    \vec{v}_b\cdot{\nabla} G +\frac{4}{3} G\,{\nabla}\cdot\vec{v}_b=0\,,
\end{align}
with the explicit expression
\begin{align}
    G(s_{\vec{r}}) \propto \exp \left[ - \int_{s_0}^{s_{\vec{r}}} \frac{4}{3} \frac{\nabla \cdot \vec{v}_b(s)}{v_b(s)} \d s \right] = \exp \left[- \int_{s_0}^{s_{\vec{r}}} \frac{4}{3} \left(\nabla \cdot \hat{v}_b(s) + \frac{1}{v_b(s)}\frac{\d v_b}{\d s}(s) \right) \d s \right]\,.
    \label{eq:F_equation}
\end{align}
The factor $G(s_{\vec{r}})\, G^{-1}(s_{i})$ in Eq.~(\ref{eq:f_solution}) accounts for the effects of flow compression and expansion. As shown in Eq.~(\ref{eq:F_equation}), $G$ contains two distinct contributions: the first term arises from the convergence or divergence of $\hat{v}_b$, which is aligned with the magnetic field lines, while the second term captures the influence of bulk velocity variations on particle density. In the case of a spherically symmetric magnetic field and bulk velocity similar to free-fall, $G(s_{\vec{r}}) \propto r^{-2}$, consistent with the results in Refs.~\cite{Aloisio:2004hy,Regis:2008ij}.

Notice that the above discussion is not in fully general relativistic setup. We will discuss the unceratinties arising from this treatment in the last section, which demonstrates that the approximation we employed is robust for our constraints.

\subsection{Dark Matter Spike Annihilation}

This section examines the DM annihilation source function $q(\vec{r},p)$, which is proportional to the square of the DM energy density $\rho(\vec{r}\,)$:
\begin{equation}
q(\vec{r},p) =\frac{1}{4\pi p^2} \frac{\langle\sigma v\rangle\,\rho^2(\vec{r}\,)}{2 m^2_{\rm DM}} \sum_i {\rm BR}_i\frac{\d N^{\rm inj}_{e^\pm, i}}{\d p}(p).
\label{eq:source}
\end{equation}
Here, $m_{\rm DM}$ is the DM particle mass, and $\langle\sigma v\rangle$ denotes the thermally averaged annihilation cross-section times the relative velocity. $\d N^{\rm inj}_{e^\pm,i}/\d p$ represents the electron-positron ($e^\pm$) injection spectrum for the annihilation channel indexed by $i$, weighted by its branching ratio ${\rm BR}_i$. We focus on two benchmark annihilation channels,  $i=b\bar{b}$ and $e^{+}e^{-}$, assuming a $100\%$ branching ratio for each. The injection spectra for these channels are obtained from the {\tt PPPC}~\cite{Cirelli:2010xx} and {\tt MadDM}~\cite{Ambrogi:2018jqj} packages, with the latter specifically applied to the $e^{+}e^{-}$ channel for DM masses below $5$\,GeV.

For the DM energy density profile, we adopt the spike density model proposed in Ref.~\cite{Gondolo:1999ef} as an illustrative example. Our analysis specifically focuses on the SMBH M87$^*$, with a mass of $M_{\rm BH}= 6.5 \times 10^9\,M_{\odot}$~\cite{EventHorizonTelescope:2019dse,EventHorizonTelescope:2019ggy} and a gravitational radius of $r_g \equiv G M_{\rm BH} \approx 3.1 \times 10^{-4}\,{\rm pc}$. We follow the parameterization and normalization approach outlined in Refs.~\cite{Lacroix:2015lxa,Lacroix:2016qpq}.

Starting with a basic power-law halo profile, $\rho_{\rm halo}(r) \propto r^{-1}$, which approximates the inner regions of a Navarro-Frenk-White (NFW) profile~\cite{Navarro:1995iw}, the resulting spike profile—formed due to adiabatic accretion by the SMBH at the galactic center—is given by:
\begin{align}
\rho (\vec{r}\,)=\left\{\begin{array}{ll}
0 & \qquad r< r_{\rm crit}(\theta,a_J)\,,\\
\rho_{\rm sat} \equiv m_{\rm DM}/(\langle\sigma v\rangle\, t_{\rm BH}) & \qquad r_{\rm crit}(\theta,a_J) \leq r < r_{\rm sat}\,, \\
\rho_{\rm sp}(r) \equiv \rho_0 \,(r/r_0)^{-\gamma_{\rm sp}} & \qquad r_{\rm sat} \leq r < r_{\rm sp}\,,\\
\rho_{\rm halo}(r)= \rho_0 (r_{\rm sp}/r_0)^{-\gamma_{\rm sp}} (r/r_{\rm sp})^{-1} & \qquad r \geq r_{\rm sp}\,.
\end{array}\right.
\label{eq:DM profile}
\end{align}
The parameters in this equation are explained below:
\begin{itemize}
\item 
The parameters $\gamma_{\rm sp}=7/3$ and $r_{\rm sp}= 0.001 M_{\rm BH}^{3/2}\rho_0^{-3/2}r_0^{-7/2}$  represent the slope index and the extent of the DM spike, respectively, which are determined by the initial halo profile in the absence of the SMBH~\cite{Gondolo:1999ef}. Here, we assume an NFW halo profile, given by $\rho_{\rm halo}(r) \propto r^{-1}$~\cite{Navarro:1995iw}. The normalization factors are chosen as $\rho_0 = 3.3 \times 10^4 \, {\rm GeV\,cm^{-3}}$ and $r_0=10^5\,r_g$ to ensure that the total enclosed DM mass within $r_{\rm sp}$ is $\Delta M_{\rm BH} \approx 5\times 10^8 M_{\odot}$. This normalization approach follows Refs.~\cite{Lacroix:2015lxa,Lacroix:2016qpq}, incorporating approximately $10\%$ uncertainty in the BH mass measurement, which arises from joint analyses of stellar motion~\cite{Gebhardt_2011} and the size of the photon ring~\cite{EventHorizonTelescope:2019dse,EventHorizonTelescope:2019ggy}.

\item  
The saturation region, characterized by a density of $\rho_{\rm sat} \equiv m_{\rm DM}/(\langle\sigma v\rangle\, t_{\rm BH})$, arises due to DM annihilation~\cite{Gondolo:1999ef}. Here, we assume a BH age of $t_{\rm BH} = 10^8\,{\rm yr}$~\cite{Lacroix:2016qpq}. The radius of this region, given by $r_{\rm sat} = r_0\,(\rho_0/\rho_{\rm sat})^{1/\gamma_{\rm sp}}$, is determined to ensure a smooth transition in the density profile. For M87$^*$, the values are:
    \begin{align}
        \rho_{\rm sat} &\approx 3 \times 10^{13} \,{\rm GeV\,cm^{-3}}\left(\frac{m_{\rm DM}}{10 \, {\rm GeV}}\right) \left(\frac{\langle\sigma v\rangle}{10^{-28} \, {\rm cm^3\,s^{-1}}} \right)^{-1}\,,\\
        r_{\rm sat} &\approx 14 \, r_g  \left(\frac{m_{\rm DM}}{10 \,  \mathrm{GeV}}\right)^{-3/7} \left(\frac{\langle\sigma v\rangle}{10^{-28} \, {\rm cm^3\,s^{-1}}}\right)^{3/7}\,.\label{eq:rsat}
    \end{align}
This relation indicates that a larger annihilation cross-section leads to a more extended saturation region. If $r_{\rm sat}$ is smaller than the critical radius $r_{\rm crit}(\theta,a_J)$, the impact of annihilation on the DM profile becomes negligible.

\item The parameter $r_{\rm crit}(\theta,a_J)$ defines the inner cutoff boundary of the DM density profile. For massive particles bound by the BH gravitational potential, any orbit that dips below $r_{\rm crit}$ inevitably plunges into the BH. The dimensionless quantity $\mathcal{R}_{\rm crit} \equiv r_{\rm crit}/r_g$ is determined as the largest real root of the equation~\cite{Hod:2013mgr}:
\begin{equation}
    \mathcal{R}_{\rm crit}^4 - 4 \mathcal{R}_{\rm crit}^3 - a_J^2 (1 - 3 \cos^2 \theta) \mathcal{R}_{\rm crit}^2 + a_J^4 \cos^2 \theta + 4 a_J \sqrt{(1-\cos^2 \theta)(\mathcal{R}_{\rm crit}^5 - a_J^2 \mathcal{R}_{\rm crit}^3 \cos^2 \theta)} = 0
\end{equation}
which is non-spherical and depends on both the polar angle $\theta$ and the BH spin parameter $a_J$.

For a BH with $a_J = 0.9375$ ($0.5$), the critical radius $r_{\rm crit}$ ranges from $1.5$ ($2.9$)\,$r_g$ at the equatorial plane ($\theta = \pi/2$) to $3.5$ ($3.9$)\,$r_g$ along the polar axis $\theta = 0$. A full general relativistic analysis predicts a denser spike profile~\cite{Sadeghian:2013laa,Ferrer:2017xwm,Zhang:2024hrq} than the initial model~\cite{Gondolo:1999ef} near the cutoff radius $r_{\rm crit}$, reinforcing our benchmark choice as a conservative approach for setting constraints on DM annihilation.
\end{itemize}
\begin{figure}[h]
    \centering   \includegraphics[width=0.6\textwidth]{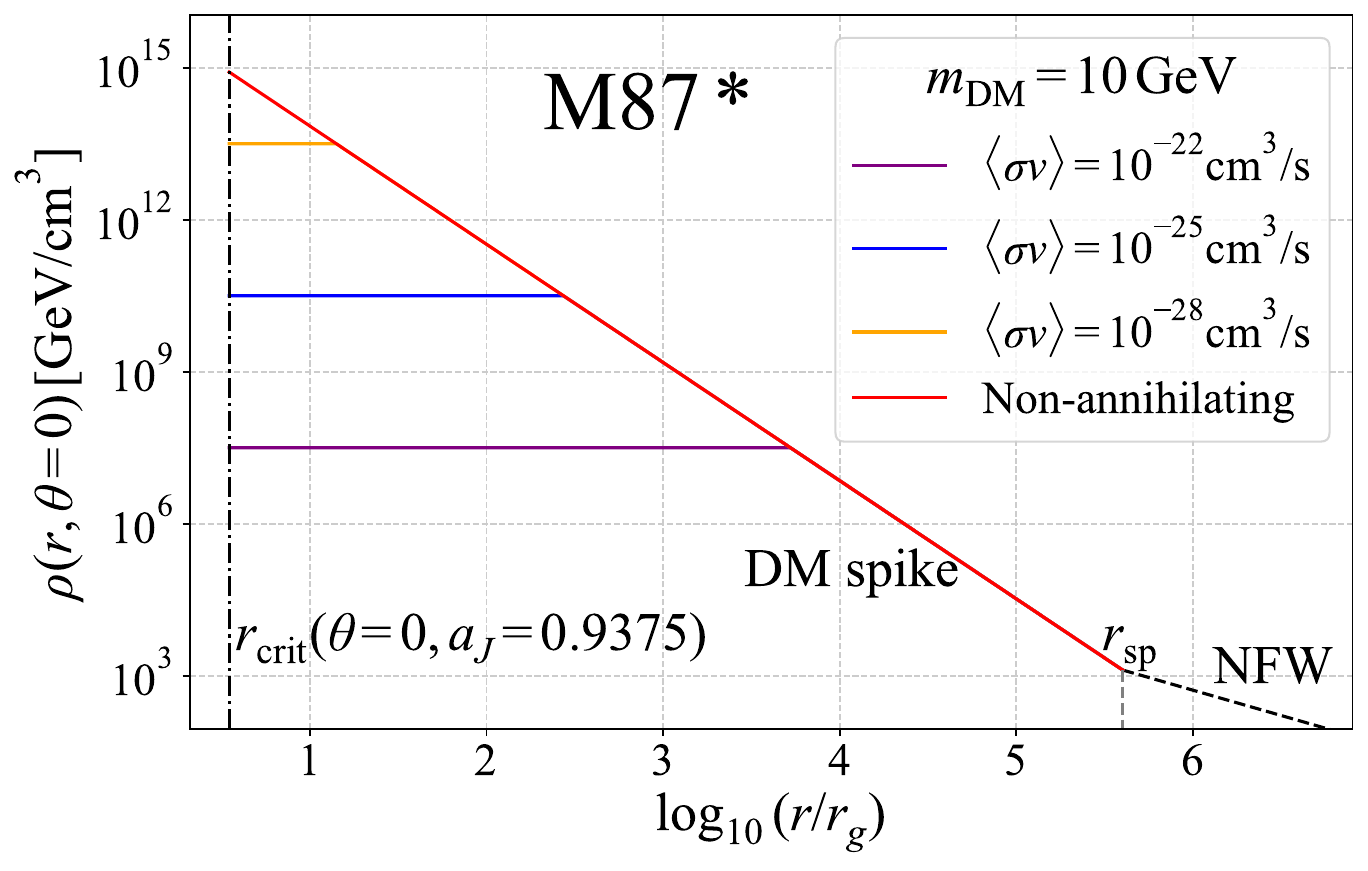}
    \caption{DM density profile outside the SMBH M87$^*$, with a gravitational radius of $r_g = 3.1 \times 10^{-4}$\,pc, as adopted in this study. The profile follows the DM spike model~\cite{Gondolo:1999ef}, transitioning from saturation at $\rho_{\rm sat}$ to an outer power-law behavior described by the NFW profile~\cite{Navarro:1995iw}, ensuring a smooth density distribution. The profiles are shown for different values of the thermally averaged annihilation cross-section $\langle \sigma v \rangle$, while fixing the DM particle mass at $10\,\mathrm{GeV}$. A larger annihilation cross-section or lower DM mass results in a lower saturation density in the inner region. The inner cutoff radius, $r_{\rm crit}$, is determined by the BH spin parameter and polar angle, set here to $a_J = 0.9375$ and $\theta = 0$, yielding $r_{\rm crit} \approx 3.5\,r_g$.}
    \label{fig:DMprofile}
\end{figure}

The benchmark DM profile outside the SMBH M87$^*$ ($r_g = 3.1 \times 10^{-4}$\,pc) is shown in Fig.~\ref{fig:DMprofile}, where we adopt the cutoff radius $r_{\rm crit}$ at the polar axis $(\theta=0)$ for a BH spin of $a_J = 0.9375$.

\subsection{Magnetic Field Profiles from GRMHD Simulations}\label{sec:BfGRMHD}

The magnetic field configuration incorporated into the transport equation is adapted from the results of general relativistic magnetohydrodynamic (GRMHD) simulations for the magnetically arrested disk (MAD) model~\cite{Mizuno:2021esc,Igumenshchev:2003rt,Narayan:2003by,McKinney:2012vh,Tchekhovskoy2015}, as performed using the \texttt{BHAC} code~\cite{Porth:2016rfi,Olivares:2019dsc}. The GRMHD-generated magnetic field profile is described in cylindrical coordinates, where the $z$-axis aligns with the polar axis of the Kerr BH, and can be converted from the Boyer–Lindquist coordinates as $(x,\phi,z)=(r \sin \theta,\phi,r \cos \theta)$. Within this framework, the magnetic field vector is formulated as:
\begin{align}
    \vec{B} = B_{x} \hat{e}_{x} + B_{\phi} \hat{e}_{\phi} + B_{z} \hat{e}_{z},
\end{align}
with $(\hat{e}_{x},\hat{e}_{\phi},\hat{e}_{z})$ representing the unit vectors in the corresponding directional coordinates. To facilitate analysis, we define dimensionless variables:
\begin{align}
\mathcal{R} \equiv \sqrt{\mathcal{X}^2+\mathcal{Z}^2} \equiv \frac{r}{r_g} , \qquad \mathcal{B} \equiv \frac{B}{1\,G}.
\end{align}

In our analysis, we consider both $a_J = 0.9375$ and $a_J = 0.5$ to account for uncertainties in the BH spin (see Sec.~\ref{sec:Uncertainties}). The GRMHD profiles are normalized such that the total synchrotron radiation intensity at $230$\,GHz matches the observed value of $0.6$\,Jy. Assuming that the time-averaged magnetic field exhibits approximate rotational symmetry, we compute the azimuthal and temporal averages of the magnetic field components using GRMHD simulation data.

We divide the magnetic field fitting procedure into two main regions: the jet region, defined by magnetic field lines that thread the event horizon near the polar axis, and the disk region, where the field lines terminate at the equatorial plane of the BH. The fitting methodology for each region is described separately below.

\begin{itemize}
\item In the jet region, we adopt a general analytic jet model based on Refs.~\cite{Pu:2017akw,Pu:2020pky} and numerically refit the magnetic field strengths using GRMHD simulation data. The magnetic field configuration in this region is predominantly parabolic, and the model employs parabolic streamlines centered around the polar axis. A self-similarity variable $\Psi \equiv \mathcal{R}(1-\cos\theta)$ is introduced to label magnetic field lines through the magnetic flux $\Phi(\Psi)$ and the field line angular velocity $\Omega_{F}(\Psi)$.

By applying Maxwell's equations within the ideal MHD framework, the magnetic field components can be depicted as:
\be\begin{split}
&
\mathcal{B}_{x} = \mathcal{X} \Phi^{\prime}/[2\pi \mathcal{R}(\mathcal{Z}+\mathcal{R})],\\
&\mathcal{B}_{\phi} = \Omega_{F}g_{\phi\phi}\Phi^{\prime}/(2 \pi \mathcal{R}),\\
&\mathcal{B}_{z} = \Phi^{\prime}/(2 \pi \mathcal{R}),
\label{eq:Bjet}
\end{split}\ee
where $\Phi^{\prime} \equiv \d \Phi/ \d \Psi$, and $g_{\phi\phi}$ is a component of the Kerr metric of the BH. 

The angular velocity of the magnetic field lines, $\Omega_F$, is given by~\cite{Pu:2020pky}:
\begin{align}
\Omega_{F}(\Psi,a_J) = \frac{\sin^{2}\theta_{H}(1+\ln \mathcal{G})}{4\,\ln 2+\sin^{2}\theta_{H}+(\sin^{2}\theta_H-2\mathcal{G})\ln\mathcal{G}}\Omega_{H}.
\label{eq:Bjet_Omega}
\end{align}
Here, $\Omega_H = a_J/(2\mathcal{R}_{H})$ denotes the angular velocity of the BH outer horizon and $\mathcal{R}_{H} = 1 + (1 - a_J^2)^{1/2}$. 
The factor $\mathcal{G} = 1 + \cos \theta_{H}$ depends on the polar angle $\theta_{H}$ of the magnetic field at the horizon, defined via $\theta_{H}(\Psi) \equiv \arccos(1 - \Psi/\mathcal{R}_H)$. The jet region corresponds to the range $0 \leq \Psi < \mathcal{R}_H$, with its boundary given by $\mathcal{R}_H = \mathcal{R}(1 - \cos\theta)$.

To match the magnetic field strength from GRMHD simulations, we use an ansatz for $\Phi^{\prime}(\Psi)$:
\begin{align}
\Phi^{\prime}(\Psi)=c_1 \exp(-c_2 \cdot\Psi),
\end{align}
yielding the best-fit coefficients:
\begin{equation}        \left(c_1,\, c_2\right) =  
        \begin{cases}     \left(123,\,0.679\right), \quad & a_{J}=0.9375, \\ \left(126,\,0.345\right),\quad & a_{J}=0.5.
        \end{cases}
\end{equation}

\item  
In the disk region, the magnetic field lines within the $x$–$z$ plane exhibit a hyperbolic-like structure. This motivates the following parameterization:
\begin{equation}
\frac{\mathcal{B}_x}{\mathcal{B}_z} = \frac{\mathcal{Z} \cdot K(\mathcal{R})}{\mathcal{X}}, \label{eq:Bdisk_sz}
\end{equation}
where $K(\mathcal{R})$ accounts for distortions near the event horizon and is modeled as a polynomial: $K(\mathcal{R}) = 1000\, \mathcal{R}^{-4} + 100\, \mathcal{R}^{-3} + 8 \,\mathcal{R}^{-2} + 4 \,\mathcal{R}^{-1} + 1$. This formulation describes two branches of hyperbolas, one centered on the $x$-axis and the other on the $z$-axis. In our analysis, which focuses on the disk region, we retain only the branch centered on the $x$-axis. The separation between the two branches defines the boundary of the disk region, determined by solving Eq.~(\ref{eq:Bdisk_sz}) under the constraint that the hyperbolic solution intersects the origin of the $x$–$z$ plane.

The overall magnetic field strength and its azimuthal component are fit using:
\be\begin{split}
\mathcal{B}_{2D} &\equiv \sqrt{\mathcal{B}_{x}^2+\mathcal{B}_{z}^2} = b_1 \mathcal{R}^{b_2 \sin(4\theta - b_3) - b_4},\\
\mathcal{B}_{\phi}&= a_1 \mathcal{R}^{a_2 \sin(4\theta - a_3) - a_4}. \label{eq:Bdisk_B}
\end{split}\ee
The best-fit parameters for the two spin cases are:
\begin{equation}        \left(a_1,\,a_2,\,a_3,\,a_4\right) =  
        \begin{cases}     \left(38.1,\,0.511,\,2.34,\,1.81\right), \quad & a_{J}=0.9375, \\ \left(19.4,\,0.395,\,2.72,\,1.56\right),\quad & a_{J}=0.5.
        \end{cases}
\end{equation}
\begin{equation}        \left(b_1,\,b_2,\,b_3,\,b_4\right) =  
        \begin{cases}     \left(52.1,\,0.579,\,2.39,\,2.29\right), \quad & a_{J}=0.9375, \\ \left(94.1,\,0.596,\,2.70,\,2.39\right),\quad & a_{J}=0.5.
        \end{cases}
\end{equation}
As expected, the fraction $\mathcal{B}_{\phi}/\mathcal{B}$ increases with BH spin.

\end{itemize}

\subsection{Bulk Velocity of Electron-Positron Flows}\label{sec:BVF}

Analogous to the magnetic field configuration discussed in the previous subsection, the bulk velocity $\vec{v}_b$ exhibits distinct behaviors in the jet and disk regions, primarily due to the rotation of magnetic field lines in the jet.

In the disk region, the SMBH’s gravitational pull drives electrons and positrons inward along magnetic field lines. As a result, the direction of the bulk velocity aligns with the field lines, with a radial component consistently directed toward the BH. Since the Lorentz force does not change the kinetic energy of a charged particle, the magnitude of the bulk velocity is predominantly governed by gravity and approximated by the radial infall velocity $v_b=(2\,r_g/r)^{1/2}$.

In the jet region, magnetic field lines extend to the event horizon and co-rotate with the BH due to frame dragging. This rotation imparts angular momentum to the plasma, inducing rotational motion around the BH’s polar axis with angular velocity $\vec{\Omega}$. In the co-rotating frame, the electron-positron dynamics are influenced by both the BH’s gravity and the inertial centrifugal potential~\cite{Pu:2017akw,Pu:2020pky}. Assuming a constant angular velocity $\vec{\Omega} \equiv \Omega_F(\Psi) \hat{e}_z$ along each magnetic field line, as given in Eq.~(\ref{eq:Bjet_Omega}), the bulk velocity $\vec{v}_b$ is determined from energy conservation:
\begin{equation}
    \frac{\d \gamma_{v_b}}{\d s} =\hat{v}_b(\vec{r}\,) 
    \cdot \left( -\frac{r_g\gamma_{v_b}}{r^3} \vec{r}  - \gamma_{v_b}\vec{\Omega} \times \left( \vec{\Omega} \times \vec{r} \right) \right)\,,\label{eq:gamma_v}
\end{equation}
where $\gamma_{v_b} = (1-v_b^2)^{-1/2}$ is the Lorentz factor associated with the bulk motion.

The stagnation surface—where $\mathrm{d} \gamma_{v_b}/\mathrm{d} s = 0$—marks the transition between inflow and outflow, and is given by:
\begin{equation}
    r_{\rm stag} (\Psi) = \frac{r_g}{\Omega_F(\Psi) \sqrt{\Psi}}\,,
\end{equation}
such that the bulk velocity vanishes, $\vec{v}_b = 0$, at this location. 

By defining an effective potential:
\begin{align}
V_{\rm eff}=-\frac{r_g}{r}-\frac{1}{2}\Omega^2 r^2\sin^2\theta\,,
\end{align}
Eq.~(\ref{eq:gamma_v}) simplifies to: $\d \ln\gamma_{v_b}=-\d V_{\rm eff}(r,\theta)$. Integrating this relation yields the bulk velocity magnitude on either side of the stagnation surface:
\begin{equation}
    v_b=\sqrt{1-e^{2[V_{\rm eff}(r,\theta)-V_{\rm eff}(r_{\rm stag},\theta_{\rm stag})]}}\,,
\label{eq:v_b}
\end{equation}
where $(r_{\rm stag},\theta_{\rm stag})$ corresponds to the point at which the magnetic field line passing through $(r, \theta)$ intersects the stagnation surface.

\subsection{Numerical Procedure and Results}

We discuss the detailed procedure for calculating the electron-positron phase space, $f_e(\vec{r},p)$, which is a solution to the propagation equation in Eq.~(\ref{eq:expanded}). An integral solution is provided in Eq.~(\ref{eq:f_solution}), integrating the injection of electron/positron from DM annihilation along each streamline parameterized using $\vec{r}\,(s)$.

We first solve for the injected momentum, $p_{\rm inj}(s_{\rm inj})$, along the streamline that ultimately contributes to the momentum $p$ in $f_e(\vec{r},p)$. To expedite the calculation, we note the existence of an integral form of the solution for Eq.~(\ref{eq:dp_ds}), given by:
\begin{align}
\frac{1}{p_{\rm inj}(s_{\rm inj}) }=\chi(s_{\rm inj}, s_{\vec{r}})\left[\frac{1}{p}-\beta(s_{\rm inj}, s_{\vec{r}})\right]\,,
\label{eq:pinjeq}
\end{align}
where we introduce
\begin{align}
\chi(s_{\rm inj}, s_{\vec{r}}) &\equiv \exp \left[ \int_{s_{\vec{r}}}^{s_{\rm inj}} \frac{ \nabla \cdot \vec{v}_b(s')}{3 v_{b}(s')} \,\d s^{\prime} \right]= \left(\frac{G (s_{\vec{r}} )}{G ( s_{\rm inj})}
\right)^{1/4},\label{eq:chi}\\
\beta(s_{\rm inj},s_{\vec{r}}) &\equiv \frac{e^4}{36 \pi^2 m_e^4} \int_{s_{\vec{r}}}^{s_{\rm inj}}\,\frac{ B(s')^2 }{v_{b} (s')}\exp \left[-\int_{s_{\vec{r}}}^{s^{\,\prime}} \frac{\nabla \cdot \vec{v}_b (s'')}{3 \, v_{b}(s'')} \,\d s^{\,\prime\prime} \right] \d s^{\,\prime}\,, \label{eq:beta}
\end{align}
satisfying $\chi(s_{\vec{r}}, s_{\vec{r}}) = 1$ and $\beta(s_{\vec{r}}, s_{\vec{r}})=0$, respectively.

\begin{figure}[htb]
    \centering
     \includegraphics[width=0.6\textwidth]{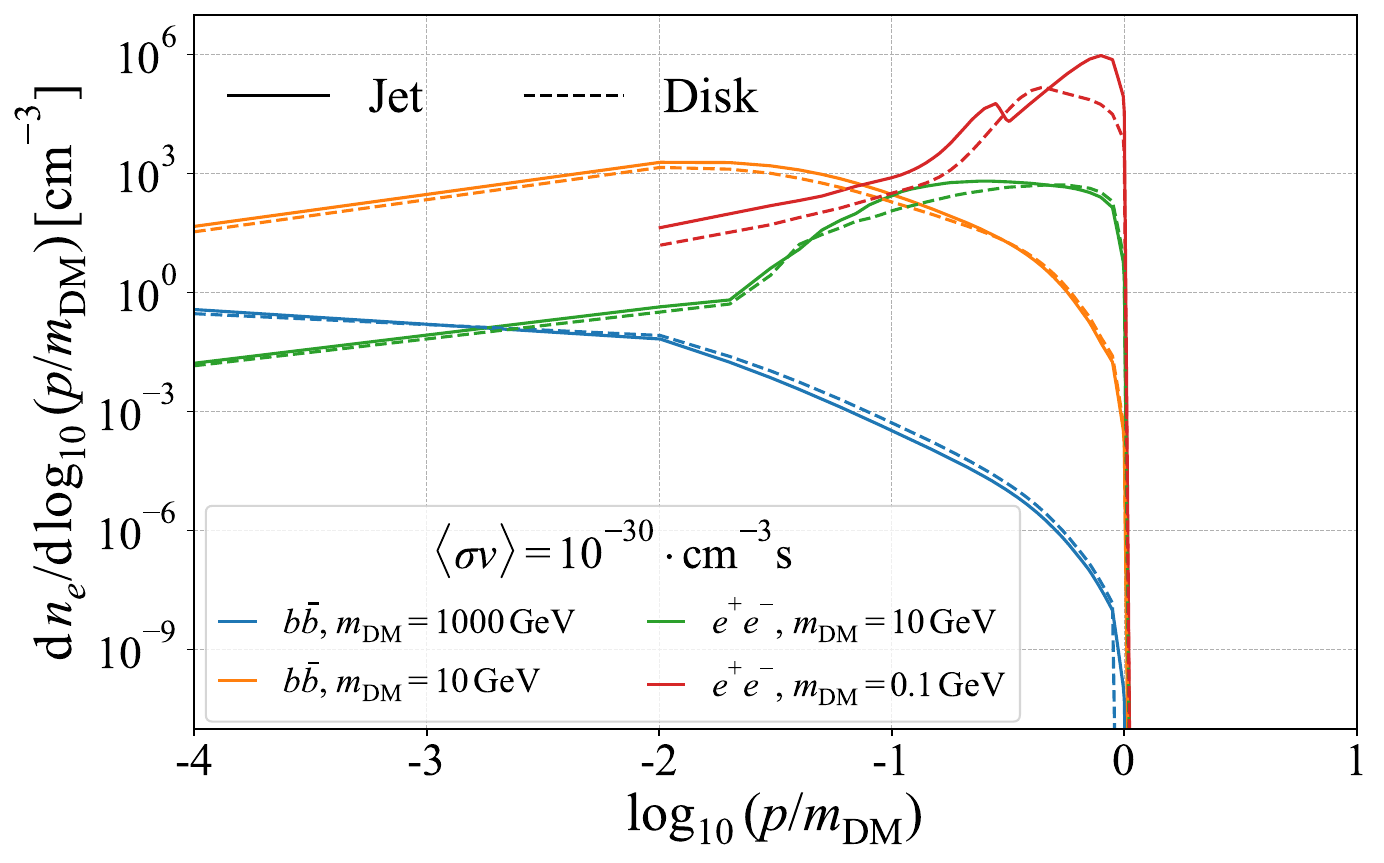}   
     \caption{Examples of the electron-positron energy spectra, $\d n_e/\d \log_{10}p\equiv 4\pi\,(\ln10)\, p^3\,f_e(\vec{r}, p)$ for both the $b\bar{b}$ and $e^{+}e^{-}$ channels, at various benchmark DM masses. Solid lines represent spectra at a typical jet location $(r,\theta) = (4\,r_g, \pi/16)$, while dashed lines correspond to a representative disk location $(r,\theta) = (4\,r_g, 7\pi/16)$. In most cases, synchrotron losses dominate over adiabatic acceleration. The $b\bar{b}$ channel yields softer spectra compared to the $e^{+}e^{-}$ channel, reflecting differences in the injection spectra. For heavier DM, stronger synchrotron cooling leads to greater spectral suppression. For $m_{\rm DM} = 0.1\,$GeV, the calculation terminates at non-relativistic momenta $p \sim \mathrm{MeV}$.}
   \label{fig:egfejetdisk}
\end{figure}

For each point $\vec{r}$, we compute the $\chi$ and $\beta$ functions along the corresponding streamline up to the boundary point $s_0$. Substituting Eqs.~(\ref{eq:pinjeq}) and (\ref{eq:source}) into Eq.~(\ref{eq:f_solution}) yields the phase space density $f_e(\vec{r},p)$. In both the jet and disk regions, our calculations are confined to the domain specified in Sec.~\ref{sec:BfGRMHD}, with $r \in [2.5\,r_g, 30\,r_g]$. For locations within this radial range but lying outside the defined jet and disk regions, we interpolate $\log_{10} f_e(\vec{r},p)$ linearly along the $\theta$ direction at fixed $r$. In the inner region where $r < 2.5\,r_g$, we set $f_e(\vec{r},p)$ equal to its value at $r = 2.5\,r_g$ for the corresponding $\theta$.

The range for $p_{\rm inj}$ is specified by the injection spectrum $\d N^{\rm inj}_{e^\pm,i}/\d E\,(p_{\rm inj})$, generated using the {\tt PPPC}~\cite{Cirelli:2010xx} and {\tt MadDM}~\cite{Ambrogi:2018jqj} packages, with the latter applied exclusively to the $e^{+}e^{-}$ channel for DM masses below $5$\,GeV. Furthermore, the injection spectrum for the $e^{+}e^{-}$ channel exhibits a sharp peak as $p_{\rm inj}$ approaches $m_{\rm DM}$. To improve numerical precision during integration, we fit the region near the peak with a half-Gaussian function, ensuring normalization to maintain the total number of injected $e^\pm$ particles. Based on the given injection spectra, the range of $p_{\rm inj}$ is determined as follows:
\begin{align}
p_{\rm inj} \in \begin{cases}
[1\,{\rm keV},\, m_{\rm DM}] & \quad \text{for}\,\, m_{\rm DM} \in [1\,{\rm MeV},\, 5\,{\rm GeV})\,,\\
[10^{-8}\,m_{\rm DM},\,m_{\rm DM}] & \quad \text{for}\,\, m_{\rm DM} \in [5\,{\rm GeV},\, 10\,{\rm TeV}] \,.
\end{cases}
\label{eq:pinj}
\end{align}
Consequently, the range of $p$ is set as $p \in [1\,{\rm MeV},\, 10\,{\rm TeV}]$, where the lower bound of $p$ is truncated at $1$ MeV to ensure that electrons and positrons remain relativistic. This broader range of $p$ compared to $p_{\rm inj}$ allows for the inclusion of both acceleration and deceleration processes that electrons and positrons undergo during propagation.

In Fig.~\ref{fig:egfejetdisk}, we present the electron-positron energy spectra $\d n_e/\d \log_{10}p\equiv 4\pi\,(\ln10)\,  p^3\,f_e(\vec{r}, p)$ for both the $e^{+}e^{-}$ and $b\bar{b}$ channels, across various benchmark DM masses. We have selected spatial positions $\vec{r}$ corresponding to two distinct points: one in the jet region $(r,\theta) = (4\,r_g, \pi/16)$ contributing to emissions in the inner shadow, and another in the disk region $(r,\theta) = (4\,r_g, 7\pi/16)$. The distribution of the electron-positron energy spectrum generally maintains the shape of the injection spectrum, although distortions and energy shifts occur due to propagation effects.

\begin{figure}[htb]
    \centering
    \includegraphics[width=0.9\textwidth]{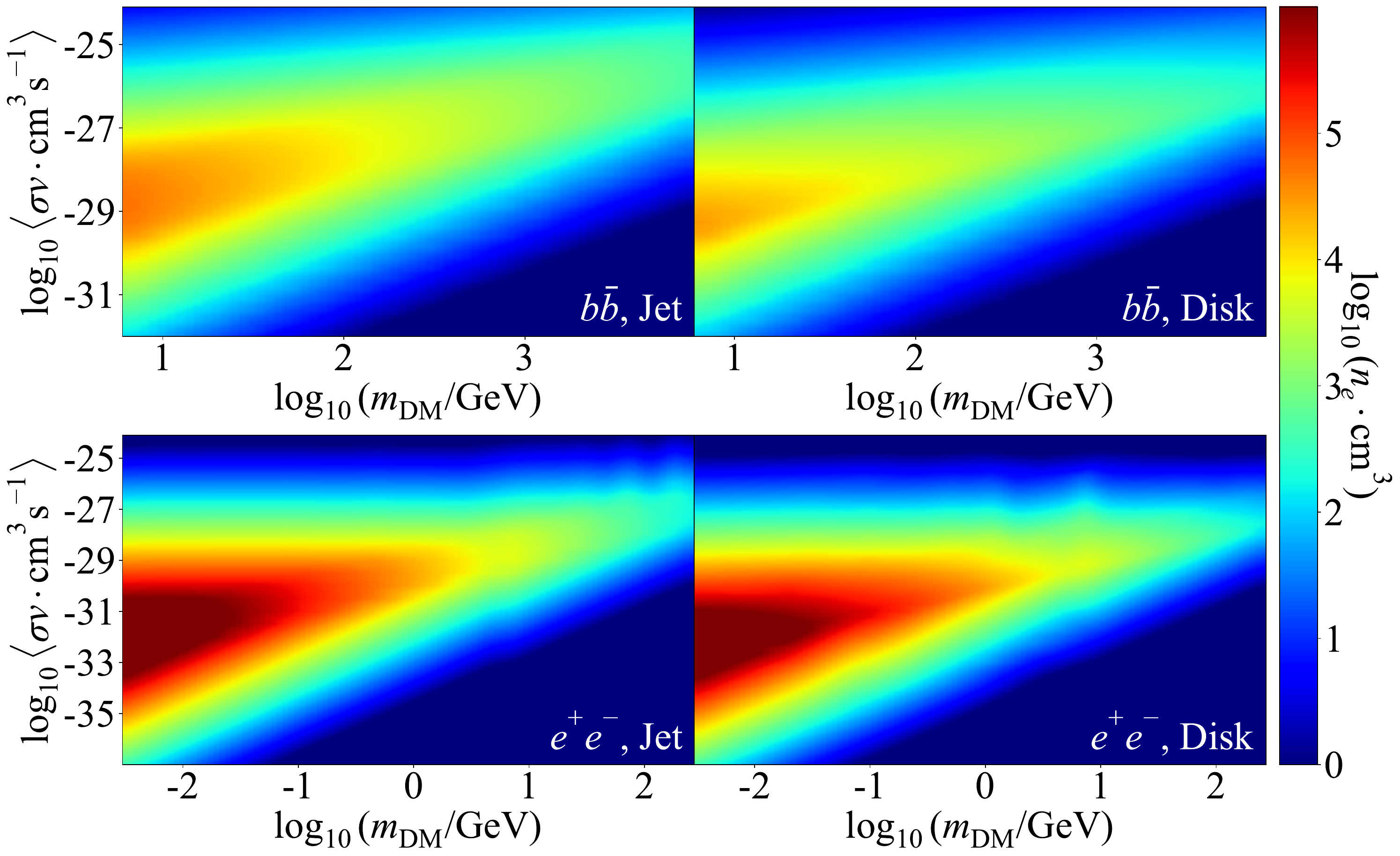}
    \caption{Electron-positron number density, $n_e(\vec{r}\,) \equiv \textstyle\int 4\pi p^2 f_e(\vec{r},p) \, \d p$, produced from DM annihilation after propagation, shown across the $(m_{\rm DM},\,\langle \sigma v \rangle)$ parameter space for both the $b\bar{b}$ and $e^{+}e^{-}$ channels. The density is evaluated at representative points in the jet region $(r,\theta) = (4\,r_g, \pi/16)$ and the disk region $(r,\theta) = (4\,r_g, 7\pi/16)$, consistent with Fig.~\ref{fig:egfejetdisk}. The horizontal contours arise from the formation of a central core in the DM profile at large cross sections, while the tilted contours reflect the steeper spike regime at lower cross sections.}
    \label{fig:neDM}
\end{figure}

In Fig.~\ref{fig:neDM}, we further illustrate the electron-positron number density, $n_e(\vec{r}\,) \equiv \textstyle\int 4\pi p^2 f_e(\vec{r},p) \, \d p$, across the $(m_{\rm DM},\,\,\langle \sigma v \rangle)$ plane. The selected locations are consistent with those used in Fig.~\ref{fig:egfejetdisk}. For both the $b\bar{b}$ and $e^+e^-$ channels, 
we observe a similar distribution pattern, which exhibits distinct characteristics in the high and low $\langle \sigma v \rangle$ regions. This delineation is attributable to the parameter space where the saturation radius of the DM profile, $r_{\rm sat}$, as defined in Eq.~(\ref{eq:rsat}), approximates the typical propagation length scale of electrons/positrons of about $10\,r_g$, represented by the relation:
\begin{align}
\langle \sigma v \rangle_{\rm sat} \approx
10^{-28}\, \left( \frac{m_{\rm DM}}{22\, \mathrm{GeV}} \right) \times \left( \frac{r_{\rm sat}}{10\, r_g} \right)^{7/3} {\rm cm^3\,s^{-1}}\,.
\end{align}
The analysis leads to two key observations:
\begin{itemize}
    \item In regions where $\langle \sigma v \rangle \geq \langle \sigma v \rangle_{\rm sat}$, the DM density relevant for propagation saturates at  $\rho(\vec{r}\,)=\rho_{\rm sat} \equiv m_{\rm DM}/(\langle\sigma v\rangle\, t_{\rm BH})$.  Consequently, $n_e(\vec{r}\,) \propto \langle\sigma v\rangle 
   \rho_{\rm sat}^2 / m_{\rm DM}^2$ scales as $1/\langle\sigma v\rangle$, independent of $m_{\rm DM}$, which corresponds to the nearly horizontal contour line in the high $\langle\sigma v\rangle$ region depicted in Fig.~\ref{fig:neDM}.
    \item When $\langle \sigma v \rangle \ll \langle \sigma v \rangle_{\rm sat}$, the nearby DM energy density conforms to the spike profile $\rho(\vec{r}\,) = \rho_0 \,(r/r_0)^{-7/3}$. Therefore,
    $n_e(\vec{r}\,)$ varies as $\langle\sigma v\rangle /m_{\rm DM}^2$, aligning with the contour slope in the low $\langle\sigma v\rangle$ region depicted in Fig.~\ref{fig:neDM}.
\end{itemize}

\section{Covariant Radiative Transfer and Intensity Map}

Following the computation of the electron-positron spectrum, $f_e(\vec{r},p)$, we employ the covariant radiative transfer formalism~\cite{Gammie_2012,Dexter:2016cdk} to determine the intensity at each point on the observer plane and compare it with the astrophysical background. This section details the methodology, utilizing the \texttt{RAPTOR} package~\cite{Bronzwaer:2018lde,Bronzwaer:2020kle} for numerical calculations.

To compute the flux, we first solve for geodesics that connect the observer to the BH using the backward ray tracing method~\cite{Bardeen:1972fi,Luminet:1979nyg}. Each point on the observer plane corresponds to a unique initial geodesic direction. The critical curve on the observer plane delineates regions where geodesics either escape to infinity or terminate at the BH~\cite{1965SvA.....8..868P,1968ApJ...151..659A,Luminet:1979nyg}.

Next, we integrate emissions along these geodesics, tracing from the BH toward the observer. This integration follows the covariant radiative transfer equation~\cite{Gammie_2012,Dexter:2016cdk}:
\begin{equation}
    \frac{\d}{\d \lambda}\left(\frac{I_\nu}{\nu^3}\right) = \frac{j_\nu}{\nu^2}-\nu \alpha_\nu\left(\frac{I_\nu}{\nu^3}\right),
\end{equation}
which is formulated in a local reference frame. Here, $\lambda$ is the affine parameter along the geodesics, $I_\nu$ is the intensity, $j_\nu$ is the emissivity, and $\alpha_\nu$ is the absorption coefficient. The subscript $\nu$ denotes the photon frequency in the local frame, which relates to the observed frequency $\nu_0$ via a redshift factor. The computation can be performed in any frame since the quantities $I_\nu/\nu^3$, $j_\nu/\nu^2$, and $\nu\alpha_\nu$ remain invariant under Lorentz transformations.

In practice, we select the plasma frame, where the momentum distribution of electrons/positrons at each spatial point is isotropic. The emissivity from synchrotron radiation in this frame is expressed as
\begin{equation}
    j_\nu = \int  f_e(\vec{r},p) \, \eta\, 4 \pi p^2 \, \d p, \qquad  \eta = \frac{\sqrt{3} e^3 B \sin\theta_B}{8\pi^2 m_e }F\left(\frac{\nu}{\nu_c}\right),\qquad F(x) \equiv x\int _x^{\infty} K_{5/3}(\zeta)\,\d \zeta,
    \label{eq:jsyn}
\end{equation}
where $\eta$ encapsulates the averaged angular power spectral density of synchrotron radiation at frequency $\nu$, emitted by a single relativistic electron with momentum $p$ orbiting in a magnetic field $\vec{B}$. $\theta_B$ denotes the pitch angle between $\vec{B}$ and the photon's spatial momentum, and $K_{5/3}(\zeta)$ represents a modified Bessel function. The critical frequency, $\nu_c \equiv 3 e B p^2 \sin{\theta_B}/(4\pi m_e^3)$, characterizes the peak of the synchrotron radiation spectrum. Similarly, the plasma frame's absorption coefficient integrates over $\eta$:
\begin{equation}
    \alpha_\nu =-\frac{1}{\nu^2} \int \frac{\d f_e(\vec{r}, p)}{\d p} \, \eta\,4\pi p^2 \d p.
\end{equation}
The plasma frame is distinguished from the BH frame by a boost transformation, employing the bulk velocity of electrons/positrons, $\vec{v}_b$. To transform the spatial velocity into a 4-velocity, $u^\mu$, we enforce $\vec{u} = u^0\,\vec{v}_b$, ensuring it meets the normalization condition $u_\mu u^\mu = -1$. With the observed frequency $\nu_0$ set at infinity, we determine the frequency at a given plasma frame as follows:
\begin{equation}
    \nu = \left|\frac{1}{2\pi}g_{\mu\beta} k^\mu u^\beta\right|.
    \label{eq:rf}
\end{equation}
Here, $g_{\mu\beta}$ represents the Kerr metric. $k^\mu = \d x^\mu/\d \lambda$ represents the photon's momentum, tracing along the geodesics, normalized by fixing the affine parameter at infinity such that $k^0 = 2\pi\nu_0$, where $x^\mu$ indicates the coordinate components within the Kerr metric. In scenarios where $\vec{v}_b=0$, the deviation from $\nu_0$ results solely from gravitational redshift, simplifying to  $\nu = \nu_0/\sqrt{|g_{00}|} = \nu_0/\sqrt{1-2GM_{\rm BH}/r}$ for a non-rotating BH. Conversely, far from the BH, the Doppler shift can predominate, with $\nu = \nu_0/\sqrt{(1+v \cos\theta_{v})/(1-v \cos\theta_{v})}$, where $\theta_v$  is the pitch angle between $\vec{v}_b$ and $\vec{k}$. For plasma surrounding an SMBH, both gravitational redshift and Doppler shift contribute.

We independently calculate the intensity map from a GRMHD profile of a MAD~\cite{Igumenshchev:2003rt,Narayan:2003by,McKinney:2012vh,Tchekhovskoy2015} and from DM annihilation. The GRMHD profile assumes a thermal electron distribution characterized by a local temperature, $T_e(\vec{r}\,)$. Under this thermal distribution, the emissivity and absorption coefficients are given by:
\begin{equation}
    j^{\text{th}}_\nu=\frac{\sqrt{3}n_e e^3 B \sin{\theta_B} }{8\pi m_e} \int_0^{\infty}  F\left(\frac{\nu}{\nu_c^{\rm th}\,z^2}\right)\,e^{-z}\,z^2\d z,\qquad \alpha^{\text{th}}_\nu=j^{\text{th}}_\nu\,\frac{\exp \left(2\pi \nu /  T_e\right)-1}{4\pi \nu^3}.
\end{equation}
In this expression, $\nu_c^{\rm th} \equiv 3 e B T_e^2 \sin{\theta_B}/(4\pi m_e^3)$, derived from $\nu_c$ by substituting momentum $p$ with temperature $T_e$. For the DM annihilation contribution, treated as a deviation from the astrophysical background, we consider the absorption coefficient, $\alpha_\nu$, to originate from the GRMHD model. It is found that absorption is consistently subdominant in a MAD profile:
\be \alpha_\nu \approx 1.3\times10^{-4} r_g^{-1}\left(\frac{n_e}{10^{4}\,\mathrm{cm}^{-3}}\right)\left(\frac{T_e}{20\,m_e}\right)^{-3}\left(\frac{\nu}{230\,\mathrm{GHz}}\right)^{-1},\ee
aligning with observations that favor an optically thin accretion flow at $\nu_0 = 230$\,GHz~\cite{Yuan:2014gma,Prieto:2015efa, EventHorizonTelescope:2019pgp}.

Two key parameters must be defined for the GRMHD profile: the overall normalization of physical quantities and the electron temperature $T_e$, since only the proton temperature $T_p$ is directly provided. The electron temperature can be inferred using the relation~\cite{Moscibrodzka:2015pda,Bronzwaer:2018lde}:
\begin{equation}
    \frac{T_{p}}{T_{e}} = R_{\text{low}} \frac{1}{1+\beta_{p}^2} + R_{\text{high}} \frac{\beta_{p}^2}{1+\beta_{p}^2},
\end{equation}
where $\beta_p \equiv P_{\text{gas}}/P_{\text{mag}}$ represents the ratio of gas pressure to magnetic pressure, with $P_{\text{mag}}$ being defined as $B^2/2$. Here, $R_{\text{high}}$ and $R_{\text{low}}$ represent the temperature ratios of protons to electrons in areas of high and low magnetic field strengths, respectively. In the maintext, we adopt fiducial values of $R_{\text{high}} = R_{\text{low}} = 1$, consistent with those supported by EHT observations~\cite{EventHorizonTelescope:2019pgp,EventHorizonTelescope:2021srq,EventHorizonTelescope:2023gtd}. In the final section, we examine potential uncertainties in these parameters based on both observational data and two-temperature GRMHD simulations, and show that their impact on our constraints is minor.

The exclusion region for the $e^+e^-$ annihilation channel displays a notable shift for DM masses below GeV scales, differing significantly from the higher mass region and the $b\bar{b}$ channel, as illustrated in Fig.~3 of the maintext. Specifically, the $b\bar{b}$ channel excludes regions of highest  $n_e$ shown in Fig.~\ref{fig:neDM}, whereas the $e^+e^-$ channel exhibits a distinct turning point around $0.1$\,GeV. This shift is attributed to a pronounced drop in the synchrotron radiation flux at $230$\,GHz for DM masses below $0.1$\,GeV. Accounting for gravitational redshift near an emission point approximately $4\,r_g$ from the SMBH, the photon frequency in the plasma frame is about $300$\,GHz. The synchrotron radiation spectrum, for an electron or positron with momentum $p$, adheres to the $F$ function as defined in Eq.~(\ref{eq:jsyn}). This function presents a gentle slope $F \sim (\nu/\nu_c)^{1/3}$ for $\nu/\nu_c \ll 1$, transitioning to a steep decline $F \sim \exp(-\nu/\nu_c)$ for $\nu$ exceeding $\nu_c$. The critical frequency $\nu_c \equiv 3 e B p^2 \sin{\theta_B}/(4\pi m_e^3)$ is estimated numerically as:
\begin{equation}
    \nu_c \approx 180 \, \mathrm{GHz} \times \left(\frac{B}{7\, \mathrm{G}}\right) \times \left(\frac{p}{0.1 \, \mathrm{GeV}}\right)^2.
\end{equation}
Referencing Fig.~\ref{fig:egfejetdisk}, the spectrum of electrons and positrons predominantly corresponds to $p \sim m_{\rm DM}$. Thus, most synchrotron radiation emanating from the annihilation products of DM with mass below $0.1$\,GeV fails to contribute significantly at $\nu = 230$ GHz.

\section{Critical Curve and Inner Shadow Contour}

In this section, we elucidate the equations defining the critical curve and the contour of the inner shadow, as depicted in Fig.~2.

We start by defining a universal coordinate system $({\alpha}, {\beta})$ on the observer plane, where the ${\alpha}$-axis is perpendicular to, and the ${\beta}$-axis is parallel with, the projected spin of the BH~\cite{Bardeen:1972fi,Luminet:1979nyg}. The impact parameters $\alpha$ and $\beta$ are dimensionless, normalized by the gravitational radius $r_g$.

Each point on this plane corresponds to a null geodesic in Kerr spacetime, described by two conserved quantities: the energy-rescaled angular momentum ${\lambda}$ and the Carter constant ${\eta}$, expressed as~\cite{Carter:1968rr,Gralla:2019ceu,Gralla:2019drh,Johnson:2019ljv,Lupsasca:2024wkp}:
\be \begin{split}
    {\lambda} &= -{\alpha} \sin \theta_o, \\ 
    {\eta} &= ({\alpha}^2-a_J^2)\cos^2 \theta_o + {\beta}^2, \label{eq:etalambda}
\end{split}\ee
with $\theta_o$ denoting the observer's inclination angle relative to the BH, which for M87$^*$ is considered to be $163^\circ$.

The critical curve delineates the boundary on the observer plane between geodesics that terminate on the BH (inner region) and those that escape to infinity (outer region). The corresponding geodesics form bound orbits around the Kerr BH~\cite{1965SvA.....8..868P,1968ApJ...151..659A,Luminet:1979nyg,Falcke:1999pj,Gralla:2019xty}, maintaining a constant radius $\mathcal{R}_c$ within the range $\mathcal{R}_c\in[\mathcal{R}_c^-, \mathcal{R}_c^+]$, where
\begin{equation}
    \mathcal{R}_{c}^{\pm} \equiv 2 \left[ 1+ \cos \left( \frac{2}{3} \arccos (\pm a_J) \right) \right].
\end{equation}
The determination of bound orbits requires identifying the double roots of the radial potential in the geodesic equation, leading to
\be \begin{split}
    {\lambda}_c &= a_J+ \frac{\mathcal{R}_c}{a_J}\left[ \mathcal{R}_c -\frac{2 \Delta(\mathcal{R}_c)}{\mathcal{R}_c-1} \right],\\
    {\eta}_c &= \frac{\mathcal{R}_c^3}{a_J^2 } \left[ \frac{4 \Delta(\mathcal{R}_c)}{(\mathcal{R}_c-1)^2}-\mathcal{R}_c\right],\label{eq:lambdaetarc}
\end{split}\ee
where $\Delta(\mathcal{R}_c) = \mathcal{R}_c^2 + a_J^2 - 2 \, \mathcal{R}_c$. By equating Eq.~(\ref{eq:etalambda}) with Eq.~(\ref{eq:lambdaetarc}), we construct a parametric representation of the critical curve in the ${\alpha}-{\beta}$ plane for different values of $\mathcal{R}_c$.

The contour of the inner shadow is delineated as the lensing image of the BH's equatorial horizon, as presented in Ref.~\cite{Chael:2021rjo}. This contour becomes measurable when the emission originates from the BH's equatorial plane and reaches out to the horizon. For observers with a nearly face-on view, where $|\cos \theta_o| \approx 1$, a fitting function has been derived to accurately depict the contour on the observer's plane~\cite{Chael:2021rjo}:
\begin{equation}
    {\rho}= \left[ 2 \sqrt{\mathcal{R}_H} + \left( 1+\frac{1}{2} \cos^2 \theta_o\right) \arctan \left[ \sin \varphi \tan \theta_o \right] \right],
\end{equation}
where $({\rho}, \varphi)$ represents the polar coordinates on the observer's plane, linked to Cartesian coordinates through the relationship $({\alpha}, {\beta}) = ({\rho}\cos\varphi,{\rho}\sin\varphi)$.

\section{Potential Uncertainties}

In this section, we examine various astrophysical and modeling uncertainties relevant to our study. For each key parameter, we discuss plausible variations based on observations and simulations, and evaluate their impact on our results. We also address uncertainties associated with our use of a non–fully general relativistic (GR) framework in the propagation equations. As we show below, our constraints remain robust against all these uncertainties.

\subsection{Astrophysical Uncertainties}\label{sec:Uncertainties}

Our analysis focuses on the SMBH M87$^*$, whose inclination angle is well constrained due to the presence of a relativistic jet, and whose observational features are well reproduced by GRMHD simulations. Observations suggest that M87$^*$ is best described by a Radiatively Inefficient Accretion Flow (RIAF), as expected for low-luminosity active galactic nuclei~\cite{Yuan:2014gma,EventHorizonTelescope:2019pgp}. Within the RIAF framework, GRMHD simulations typically adopt two classes of accretion models: the Standard and Normal Evolution (SANE) model and the MAD model. EHT observations—particularly linear polarization measurements and jet morphology—strongly favor the MAD scenario, in which dynamically dominant magnetic fields regulate the accretion flow~\cite{EventHorizonTelescope:2021srq}. This conclusion is further supported by recent analyses of circular polarization~\cite{EventHorizonTelescope:2023gtd}. Based on this evidence, our study focuses exclusively on MAD-type models.

Within the MAD framework, the primary model-dependent parameters include the BH spin $a_J$ and the plasma parameters $R_{\rm high}$ and $R_{\rm low}$, which characterize the ion-to-electron temperature ratio in regions of weak and strong magnetization, respectively. Below, we discuss plausible ranges for these parameters and assess their influence on our results.

In addition to these plasma and spin parameters, uncertainties in the mass of M87$^*$ and the normalization of the DM density profile can also affect our conclusions. We quantify these effects through scaling arguments, focusing on their impact on the electron-positron number density $n_e(\vec{r}\,)$ and the inferred annihilation cross section $\langle \sigma v \rangle$. We also briefly comment on the choice of DM annihilation channels.

\begin{itemize}
\item \textbf{Spin of M87$^*$ $(a_J)$:}
Current EHT observations favor a high spin for M87$^*$ based on the observed asymmetry in the photon ring, although precise determination is limited by angular resolution~\cite{EventHorizonTelescope:2019pgp}. Independent modeling based on jet dynamics places a lower bound of $a_J \geq 0.5$~\cite{Cruz-Osorio:2021cob}. Accordingly, we adopt $a_J = 0.9375$ as the fiducial value in the maintext and include an additional case with $a_J = 0.5$ to assess the impact of spin uncertainty.

To accommodate different spin values, we employ a general analytic jet model, described in Sec.~\ref{sec:BVF} and Refs.~\cite{Pu:2017akw,Pu:2020pky}, which allows for consistent treatment across a range of $a_J$. In Fig.~\ref{Fig:a05compare}, we present the resulting constraints on the DM annihilation cross section, based on magnetic field profiles fitted from GRMHD simulations for $a_J = 0.5$. Compared to the $a_J = 0.9375$ case, the exclusion contours shift only slightly toward higher cross sections. This shift arises because a lower spin reduces centrifugal support, leading to a larger stagnation surface and thus greater accumulation of electron-positron pairs in cored DM profiles. However, for spiky DM profiles, the flow $n_e \vec{v}_b$ is primarily determined by the local annihilation rate, the increased $v_b$ and smaller azimuthal magnetic field component ($\mathcal{B}_\phi/\mathcal{B}$) at lower spin result in a slightly lower $n_e$, thereby mildly weakening sensitivity at small cross sections.

\begin{figure}[htbp]
    \centering
  \includegraphics[width=0.48\textwidth]{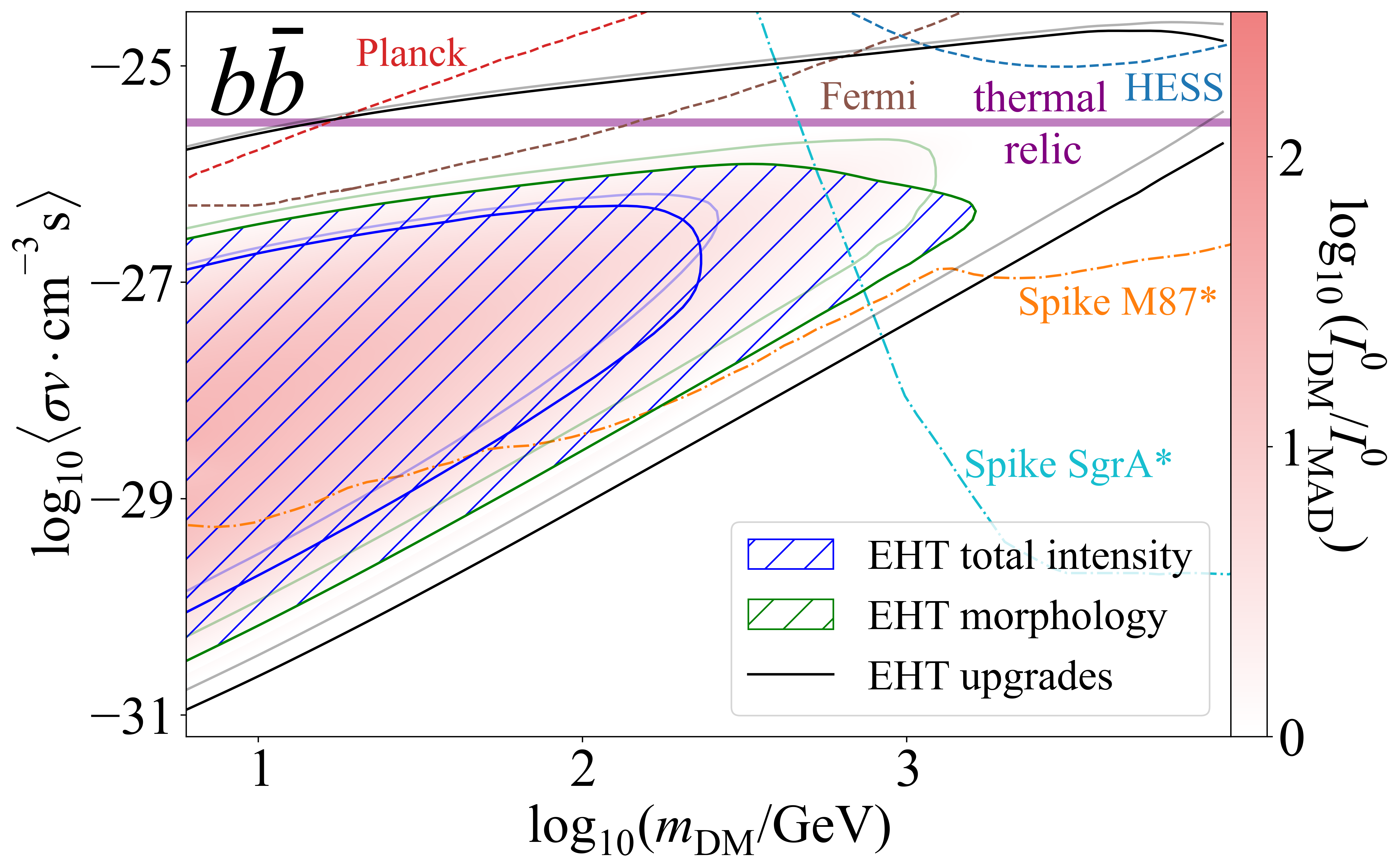}
   \includegraphics[width=0.48\textwidth]{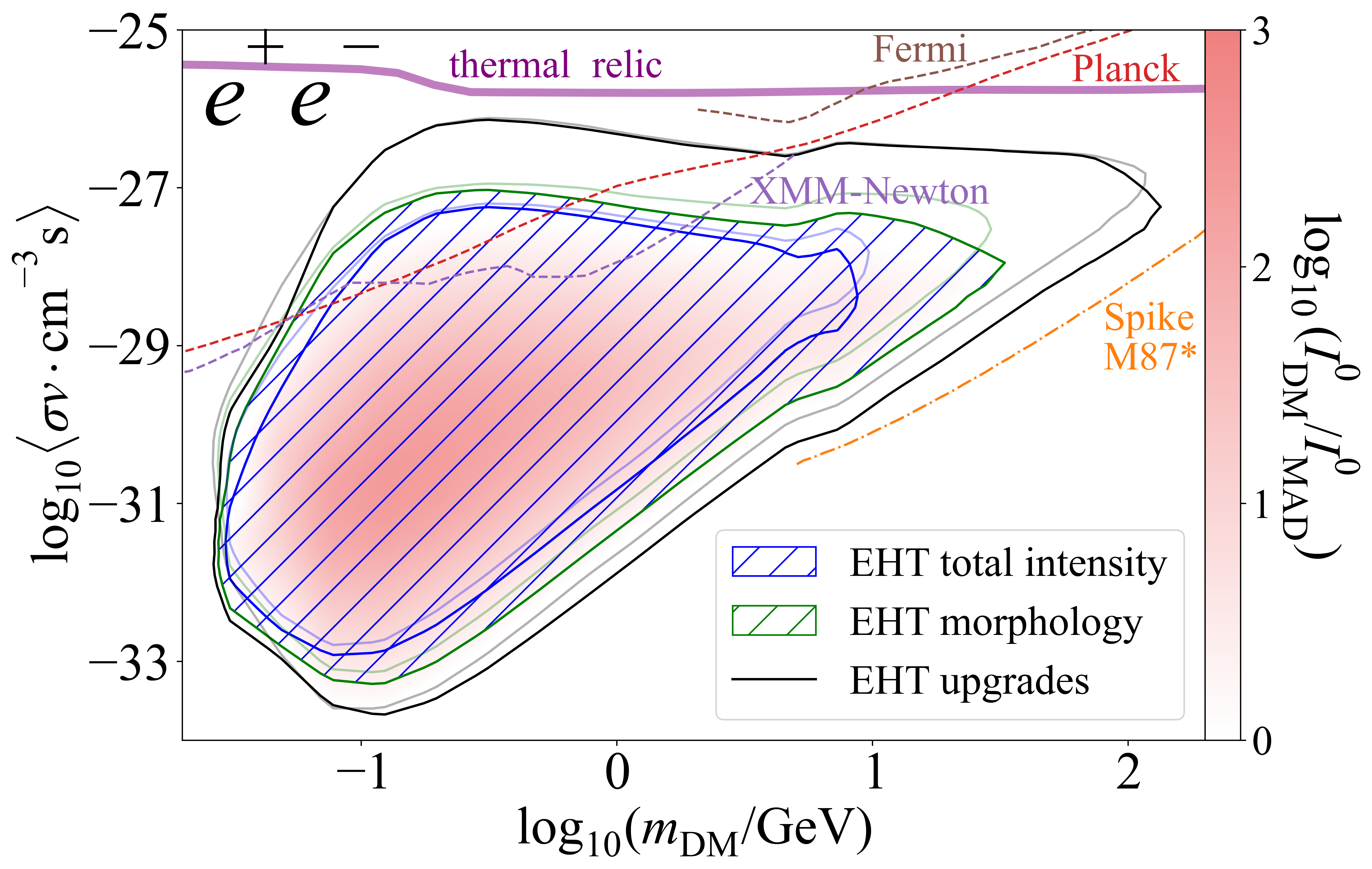}
\caption{Same as Fig.~3 in the maintext, but with additional exclusion contours (lighter colors) corresponding to a lower BH spin of $a_J = 0.5$. The exclusion contours show only slight shifts compared to the fiducial case with $a_J = 0.9375$.}
\label{Fig:a05compare}
\end{figure}

\item \textbf{$R_{\rm high}$ and $R_{\rm low}$:} 
$R_{\rm high}$ and $R_{\rm low}$ specify the proton-to-electron temperature ratio in regions of weak and strong magnetic fields, respectively. These parameters are essential for determining the electron temperature, as standard GRMHD simulations primarily yield the ion temperature. In our benchmark setup, we adopt $R_{\rm high} = R_{\rm low} = 1$, consistent with values supported by EHT observations~\cite{EventHorizonTelescope:2019pgp,EventHorizonTelescope:2021srq,EventHorizonTelescope:2023gtd}.

The plausible ranges of $R_{\rm high}$ and $R_{\rm low}$ can be inferred from both EHT data and two-temperature GRMHD simulations, the latter of which were used in Ref.~\cite{Chael:2021rjo} to model inner shadow features. In the jet region—characterized by strong magnetic fields in MAD models—electrons are preferentially heated over ions, suggesting $R_{\rm low} \lesssim 1$~\cite{Dihingia:2022aoc}. Conversely, in the disk region with weaker magnetic fields, the more rapid synchrotron cooling of electrons compared to heavier ions implies $R_{\rm high} \gtrsim 1$~\cite{Dihingia:2022aoc}. These simulations also indicate that both parameters can vary by roughly an order of magnitude. EHT observations support values of $R_{\rm low} = 1$~\cite{EventHorizonTelescope:2019pgp} and $R_{\rm low} = 1$, $10$~\cite{EventHorizonTelescope:2021srq} for MAD models, while $R_{\rm high} \lesssim 20$ is generally favored~\cite{EventHorizonTelescope:2021srq,Dihingia:2022aoc}.

To assess the impact of these uncertainties, we explore a parameter space spanning $R_{\rm low} = 0.1, 1$ and $R_{\rm high} = 1, 10, 20$, and compute the resulting intensity profiles. GRMHD configurations are normalized to match the observed EHT flux of $0.6$~Jy, which slightly affects the magnetic field strength and consequently the DM-induced synchrotron signal. We find that the intensity profiles vary only mildly with changes in $R_{\rm high}$ and $R_{\rm low}$. Since both the DM-induced and GRMHD emissions scale with magnetic field strength through synchrotron radiation, their relative behavior remains stable. To quantify the effect on exclusion sensitivity, we calculate the maximum value of $\log_{10}[I_{\rm DM}/I_{\rm MAD}]$ along the one-dimensional intensity profile (similar to the bottom panel of Fig.~2 in the maintext) for each configuration. This maximum varies by less than $36\%$ across all cases, demonstrating that our exclusion strategy is robust against variations in these plasma parameters.

 \item \textbf{Mass of M87$^*$ $(M_{\rm BH})$/gravitational radius $(r_g)$:}
    Throughout this work, all length-related quantities are expressed in units of the gravitational radius $r_g$. Accordingly, we evaluate $n_e(\vec{r}\,)$ at fixed dimensionless coordinates $\vec{r}/r_g$. The influence of the BH mass can be assessed by analyzing how the annihilation cross section $\langle \sigma v \rangle$ scales with $r_g$.

    In the low cross-section regime, the electron-positron number density from DM annihilation scales as $n_e(\vec{r}\,) \propto s_{\rm length} \langle \sigma v \rangle \rho(\vec{r}\,)^2$, where $s_{\rm length}$ is an effective propagation length scale in units of $r_g$, and $\rho(\vec{r}\,) = \rho_0 (r/r_0)^{-7/3}$. For the GRMHD-based background, the electron number density scales as $n_e(\vec{r}\,) \propto r_g^{-3}$. These scalings yield:
    \begin{equation}
        n_e (\vec{r}\,) \sim 
        \begin{cases}
            \langle \sigma v \rangle r_g^{-11/3}. \quad &\text{DM spike} \\
            r_g^{-3}. \quad &\text{GRMHD}
        \end{cases}
    \label{eq: ne_rg_relation}
    \end{equation}
Since both DM-induced and GRMHD electrons radiate via synchrotron emission in the same magnetic field configuration and are subject to the same radiative transfer process, the resulting intensity is primarily determined by the relative magnitude of $n_e$. Consequently, we get the scaling for $\langle \sigma v \rangle$ following  $\langle \sigma v \rangle \propto r_g^{2/3}$. 

The current mass estimate of M87$^*$ is $M_{\rm BH} = (6.5 \pm 0.7) \times 10^9\, M_{\odot}$~\cite{EventHorizonTelescope:2019dse,EventHorizonTelescope:2019ggy}, corresponding to a $\sim 10\%$ uncertainty. This translates into only a modest shift in the excluded range of $\langle \sigma v \rangle$, confirming the robustness of our constraints.

    \item  \textbf{DM profile normalization ($\rho_0$):}
    We next consider variations in the normalization of the DM density profile, $\rho_0$, as defined in $\rho(\vec{r}\,) = \rho_0 ~ (r/r_0)^{-7/3}$. The corresponding scaling behavior can be summarized as:
    \begin{equation}
        n_e(\vec{r}\,) \sim 
        \begin{cases}
            \langle \sigma v \rangle \rho_0^2. \quad &\text{DM spike} \\
            \text{const}. \quad &\text{GRMHD}
        \end{cases}
    \end{equation}
    Therefore, the resulting constraint on the annihilation cross section scales as $\langle \sigma v \rangle \propto \rho_0^{-2}$.

     \item \textbf{Dark matter annihilation channels:}
     Our goal is to derive model-independent constraints on the annihilation cross section for weakly interacting massive particle (WIMP) DM scenarios. Without loss of generality, we consider two representative benchmark channels: DM annihilation into $b\bar{b}$ and into $e^+e^-$ pairs. These channels capture the production of electrons and positrons through primary decays, showering, and hadronization, and represent two distinct spectral morphologies:
\begin{itemize}
    \item \textbf{$b\bar{b}$ channel:} a broad bump-like spectrum with a high yield of low-energy electrons and a cutoff at the DM mass.
    \item \textbf{$e^+e^-$ channel:} a sharp monoenergetic peak at the DM mass, along with smeared low-energy contributions from final-state radiation.
\end{itemize}
These two choices are sufficiently general, as most other annihilation channels yield electron-positron spectra that can be approximated as combinations of these two cases.
\end{itemize}

\subsection{Uncertainties Related to Approximations in Propagation Equations}

In the calculation of electron and positron propagation presented in the preceding sections, we adopt a predominantly Newtonian treatment for plasma transport. Relativistic effects are included in the kinematics of the phase space momentum $p$ (with $p \gg m_e$) and the bulk velocity $v_b$, but not in the transport dynamics. A fully general relativistic (GR) treatment would require self-consistent GRMHD simulations incorporating source injection terms, which is beyond the scope of this study. Moreover, uncertainties introduced by BH spin—particularly in the jet region where frame dragging induces rotational motion and the plasma experiences strong centrifugal forces—already lead to variations larger than those expected from neglecting GR corrections in the transport equation.

Within our framework, we employ GRMHD-derived magnetic field profiles and assume that electrons and positrons are tightly coupled to magnetic field lines due to their small Larmor radii. Consequently, GR corrections primarily affect the magnitude of the bulk velocity, while the direction is governed by the magnetic field geometry.

In the disk region, where rotational effects are negligible, we adopt $v_b = (2\,r_g/r)^{1/2}$. This expression corresponds to the full GR solution for the free-fall velocity of a particle initially at rest at infinity.

In the jet region, where the magnetic field lines are rotating, a full general relativistic calculation of $v_b$ (or $\gamma_{v_b}$) becomes significantly more complex. To estimate the GR effect, we compare the result given by Eq.~(\ref{eq:v_b}), which uses only the Newtonian gravitational potential, with the Schwarzschild solution. This comparison yields a deviation of approximately $13\%$ at $r = 4\,r_g$. The discrepancy is expected to be further reduced when accounting for the rotational effects of magnetic field lines, which exert an additional decelerating influence on the bulk flow near the BH.

A key justification for neglecting GR corrections is that the dominant emissivity arises from regions with $r > 4\,r_g$, where GR effects—typically scaling as $\mathcal{O}(r_g/r)$ relative to Newtonian estimates—remain subleading. A detailed examination of our radiative transfer calculations shows that most of the observed intensity within the inner shadow originates from regions around $4 \sim 8\,r_g$, where GR corrections are at the $\sim 10\% \sim 20\%$ level. Moreover, our analysis of electron/positron propagation reveals that particles reaching equilibrium near $4\,r_g$ are primarily injected from larger radii, where GR effects are negligible during most of their propagation. Additionally, GR effects on particle momentum are relatively weak compared to electromagnetic forces. Taken together, these considerations justify the use of a Newtonian transport framework supplemented by relativistic kinematics for phase space variables in our calculations.

\end{document}